\documentclass[twocolumn]{article}
\pdfoutput=1

\usepackage[noblocks]{authblk}
\usepackage{array}
\usepackage[square,numbers,sort&compress]{natbib}
\usepackage{tabularx}
\usepackage{graphicx}
\usepackage{hyperref}
\usepackage{gensymb}

\newcolumntype{P}[1]{>{\centering\arraybackslash}p{#1}}

\usepackage{algorithm}
\usepackage{algorithmic}

\newcommand{\angstrom}{\mbox{\normalfont\AA}}

\begin{document}

\title{DeepSurf: A surface-based deep learning approach for the prediction of ligand binding sites on proteins}
\author[1]{Stelios K. Mylonas}
\author[1]{Apostolos Axenopoulos}
\author[1]{Petros Daras}
\affil[1]{Information Technologies Institute, Centre for Research and Technology Hellas, Thessaloniki, Greece.}
\affil[ ]{$\{$smylonas@iti.gr, axenop@iti.gr$\}$}
\date{}

\maketitle

\abstract{The knowledge of potentially druggable binding sites on proteins is an important preliminary step towards the discovery of novel drugs. The computational prediction of such areas can be boosted by following the recent major advances in the deep learning field and by exploiting the increasing availability of proper data. In this paper, a novel computational method for the prediction of potential binding sites is proposed, called DeepSurf. DeepSurf combines a surface-based representation, where a number of 3D voxelized grids are placed on the protein's surface, with state-of-the-art deep learning architectures. After being trained on the large database of scPDB, DeepSurf demonstrates superior results on three diverse testing datasets, by surpassing all its main deep learning-based competitors, while attaining competitive performance to a set of traditional non-data-driven approaches. The source code of the method along with trained models are freely available at \href{https://github.com/stemylonas/DeepSurf.git}{https://github.com/stemylonas/DeepSurf.git}.}

\section{Introduction}

Structure-based drug discovery relies mostly on knowledge of potential binding sites of small compounds on protein structures. Computational binding site prediction (BSP) allows to predict in silico properties that would require much effort to establish experimentally and can enhance significantly the drug discovery process.

Through the years, a plethora of methods have been proposed for the structure-based BSP task, and, according to \cite{macari2019computational}, they can be roughly separated in three categories: the geometry-based, the energy-based and the template-based ones. Geometry-based methods (ConCavity \cite{capra2009predicting}, Fpocket \cite{le2009fpocket}, CriticalFinder \cite{dias2017multi}) predict binding cavities based solely on the geometry of the molecular surface, while energy-based methods (FTSite \cite{ngan2011ftsite}, AutoSite \cite{ravindranath2016autosite}, \cite{tsujikawa2016development}) calculate interaction energies between protein atoms and chemical probes and attempt to locate energy minima on protein's surface. On the other hand, template-based methods (Findsite \cite{brylinski2008threading}, LBias \cite{hwang2017structure}, LIBRA \cite{toti2017libra}) aim to extract binding sites on a protein by performing global or local structural alignment between this protein and a set of preexisting templates. Furthermore, consensus algorithms have been proposed that combine the results from numerous standalone methods (metaPocket2.0 \cite{zhang2011identification}, COACH \cite{yang2013protein}).

A new perspective on bioinformatics has been provided by the machine learning (ML) field. Machine learning techniques exploit the available amount of labeled data and, through the automated and iterative process of learning, manage to analyze and extract the underlying patterns that eventually correlate the data with their assigned label. Such methodologies have also been recently introduced to the structure-based BSP task \cite{jian2016predicting},\cite{krivak2018p2rank}. Specifically, Krivak and Hoksza proposed the P2Rank method \cite{krivak2018p2rank}, which employs a random-forest (RF) classifier to predict ligandability score for points placed on the solvent accessible surface of a protein. A set of chemical and geometrical features are calculated on local spherical neighbourhoods around these points and operate later as input to the RF classifier. The points receiving the highest ligandability scores are spatially clustered to finally provide the predicted binding sites.

Over the last few years, the increasing availability of large amount of data has led to the development of a subfield in ML, namely the deep learning (DL) field. DL has surpassed by far more traditional ML methods in many scientific domains (computer vision, natural language processing, etc) and has been recently applied in a variety of structural bioinformatics tasks, such as virtual screening \cite{ragoza2017protein},\cite{imrie2018protein}, binding affinity prediction \cite{stepniewska2018development},\cite{jimenez2018k} or protein structure prediction \cite{wang2016protein},\cite{senior2019protein}. DeepSite \cite{jimenez2017deepsite} was the first attempt to employ a DL architecture in structure-based BSP task, by using a rather shallow convolutional neural network (CNN) of 4 layers. DeepSite, like P2Rank, treats binding site prediction as a binary classification problem, where "binding" and "not binding" are the two considered classes. Their main difference is that DeepSite does not utilize any surface information but, on the contrary, operates on a 3D voxelized grid of the protein. For each voxel of the grid, a feature vector is computed based on the physicochemical properties of the neighboring protein atoms. Then, a sliding cuboid window of $16\times 16\times 16$ traverses the entire grid creating subgrids of features, which are then imported to the CNN. Each subgrid is finally assigned a ligandability score by the network. A very similar approach has been proposed in \cite{jiang2019novel}, where the main difference is the set of features employed. Two recently proposed methods, called Kalasanty~\cite{stepniewska2020improving} and FRSite~\cite{jiang2019frsite}, resemble to DeepSite in protein representation, since they also employ a 3D voxelization of the entire protein, but they differ on how they approach the BSP task. BSP is treated as an object-detection problem by FRSite and as a semantic segmentation problem by Kalasanty, where in both cases the desired object to be extracted is the corresponding binding site. In FRSite, a 3D version of Faster-RCNN is employed, while in Kalasanty a common segmentation architecture, called U-Net, is adapted to the needs of the specific task. According to the reported results in Kalasanty, this alternate representation has achieved higher accuracies than DeepSite.

Among the aforementioned methodologies, both DeepSite and Kalasanty exploit the inherent capabilities of deep learning architectures to learn from large databases and automatically extract features. However, the voxelized representation of the entire protein they adopt may have several limitations: It neglects any knowledge of the surface morphology, while this fixed structural discretization of the input space can lead to information loss. As stated in \cite{krivak2018p2rank}, focusing on grid points or atoms has led experimentally to significantly worse results than focusing on surface points. On the other hand, our proposed method, called DeepSurf, aims to combine effectively the learning capabilities of advanced CNN architectures with a surface-based representation of the protein  structure. This representation can exploit the binding mechanics in a more efficient way by resembling more to the actual binding process. More specifically, DeepSurf employs a 3D-CNN architecture on localized 3D grids, which are appropriately oriented and placed on a set of selected surface points.  

The main contributions of our approach are i) a new representation of the 3D protein surface is introduced, based on local voxel grids centred at sample points of the surface; ii) a novel residual network LDS-ResNet that has shown better performance than the baseline ResNet in image analysis tasks has been extended in three dimensions to be applicable to volumetric data. The proposed method has been evaluated in binding site prediction using different benchmark datasets, demonstrating superior performance among state-of-the-art approaches.

\section{Proposed method}

\subsection{DeepSurf}\label{sec:our_method}

A short outline of our method is given in Algorithm~\ref{alg:description}.

Firstly, the solvent accessible surface (SAS) of the protein is created in a triangular mesh format. The resulting mesh is usually too dense, with unnecessary redundancy of points, which can lead our algorithm to a severe computational burden. For this reason, we apply a subsequent "mesh simplification" step, where the total number of surface points is reduced by a factor of $f$ (e.g. $f=10$). This is achieved by employing the K-means clustering algorithm on the entire set of points. Our aim is to aggregate adjacent points into one cluster and, subsequently, place the local grid on a representative point of this cluster, avoiding thus redundant computations. If $n_p$ is the number of original surface points, we set the total number of created clusters to $n_p/f$. As we can see, parameter $f$ controls the density of the points to be preserved and corresponds to the average number of points per cluster. Finally, from each cluster the closest point to the cluster center is kept.

\begin{algorithm}[t]
	\caption[caption]{DeepSurf}
	\label{alg:description}
	\small
	\begin{algorithmic}[1]\normalsize
		\REQUIRE Protein structure
		% parameters ??
		\STATE Create the solvent accessible surface of the protein
		\STATE Reduce the set of surface points % "by a factor \textit{f}"
		\FOR {each point \textit{P}}
		\STATE Compute normal vector \textbf{\textit{n}} on \textit{P}
		\STATE Create local grid on \textit{P}, aligned according to \textbf{\textit{n}}
	%	\STATE Align local grid according to \textbf{\textit{n}}
		\STATE Calculate grid features 
		\STATE Import grid to 3D-CNN and get ligandability score for \textit{P}
		\ENDFOR
		\STATE Discard points with score less than $T$
		\STATE Cluster the remaining points
		\FOR {each cluster}
		\STATE Assign each cluster point to its closest protein atom 
		\STATE Form a binding site from these atoms
		\ENDFOR
		\STATE Rank binding sites by average ligandability score 
		\ENSURE Binding sites
	\end{algorithmic}
\end{algorithm}

\begin{figure}[!t]
    \includegraphics[width=0.45\textwidth]{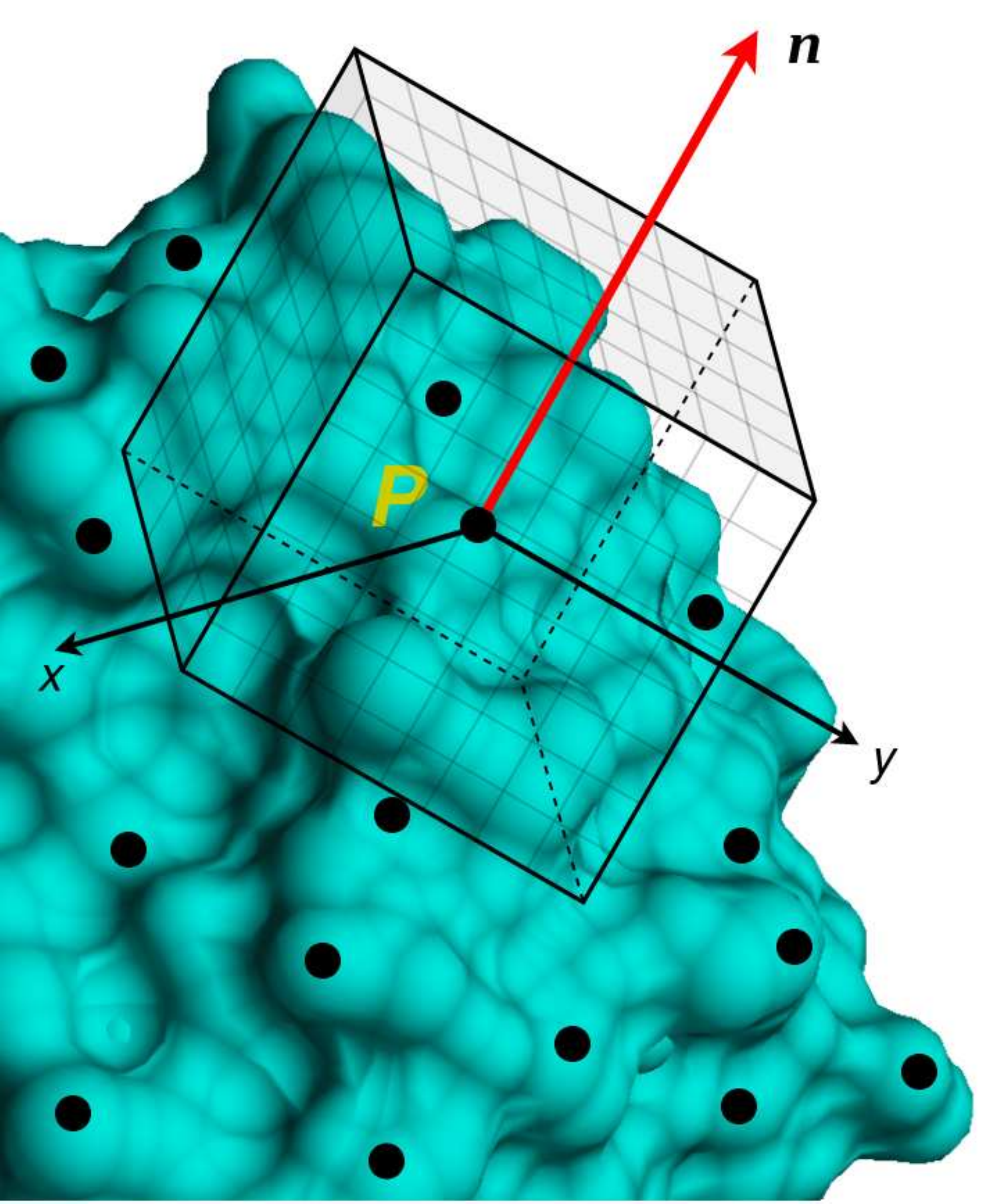}
    \caption[]{Illustration of 3D grid localized on surface point $P$ and aligned according to normal $n$.}\label{fig:3Dgrid}
\end{figure}

One issue related to the voxelized representation of a protein is the lack of rotation invariance. Specifically, due to lack of symmetry, the employed 3D cuboid grids are always rotation-sensitive and strongly depend on the arbitrary placement of the axes. Most methods attempted to address this issue by augmenting the data with random rotations during training \cite{ragoza2017protein},\cite{stepniewska2020improving}. On the other hand, P2Rank, as a non-voxelized method, bypassed this issue by utilizing symmetric spherical neighborhoods. We aim to alleviate this problem by aligning the local grids with the orientation of the normal vectors of the corresponding surface points. This alignment approach was inspired by \cite{axenopoulos2015similarity}, where local spherical regions on a protein surface were aligned according to the orientation of the normal vectors, in order to extract local shape descriptors.
An illustration of this step is shown in Fig.~\ref{fig:3Dgrid}. A local grid of size $16\times 16\times 16$ and resolution 1 \angstrom$ $ is centered on surface point \textit{P} and is oriented such that the z-axis is always parallel to the normal vector \textit{n} on \textit{P}, i.e. perpendicular to the surface. With this approach, the rotation issue is not eliminated, since random rotations are still applied during training. However, this selective initial placement of axes, instead of a random one, resulted to a more effective training and evaluation scheme. 

After the proper localization and orientation of the grid, the next step is to calculate the necessary features that will form the 4D tensor which is then imported to the 3D-CNN. We adopt here the featurization scheme initially introduced by \cite{stepniewska2018development} and used also in Kalasanty \citep{stepniewska2020improving}, which consists of 18 chemical features calculated per protein atom. Each grid voxel receives the features of the atoms inside it. The formed 4D tensor is then imported to CNN and produces at the output a ligandability score for the specific surface point. During training phase, and prior to being imported to the network, the 4D tensor is randomly rotated across one of the three axes by 90\degree. Although our approach has been tested using specific deep neural network architectures, the proposed methodology is generic, meaning that any 3D-CNN architecture that receives as input a 4D tensor and returns as output a float value in range [0,1] can be used instead. The exact network architectures employed in our experiments are elaborated in the next section. After obtaining ligandability scores for all surface points, we need to extract distinct binding sites. Points with score less than $T$ are considered not reliable and are discarded, while the remaining ones are clustered using the mean-shift algorithm \citep{comaniciu2002mean}. The main reason for selecting mean-shift instead of other clustering algorithms, is that with mean-shift we do not need to declare the number of clusters in advance. This property matches exactly to our case, since the exact number of binding sites is not known beforehand. Finally, the surface points from each cluster are assigned to their nearest protein atoms and form the desired binding sites.

\subsection{Network architectures}

\begin{figure}[!t]\centering
    \includegraphics[width=0.45\textwidth]{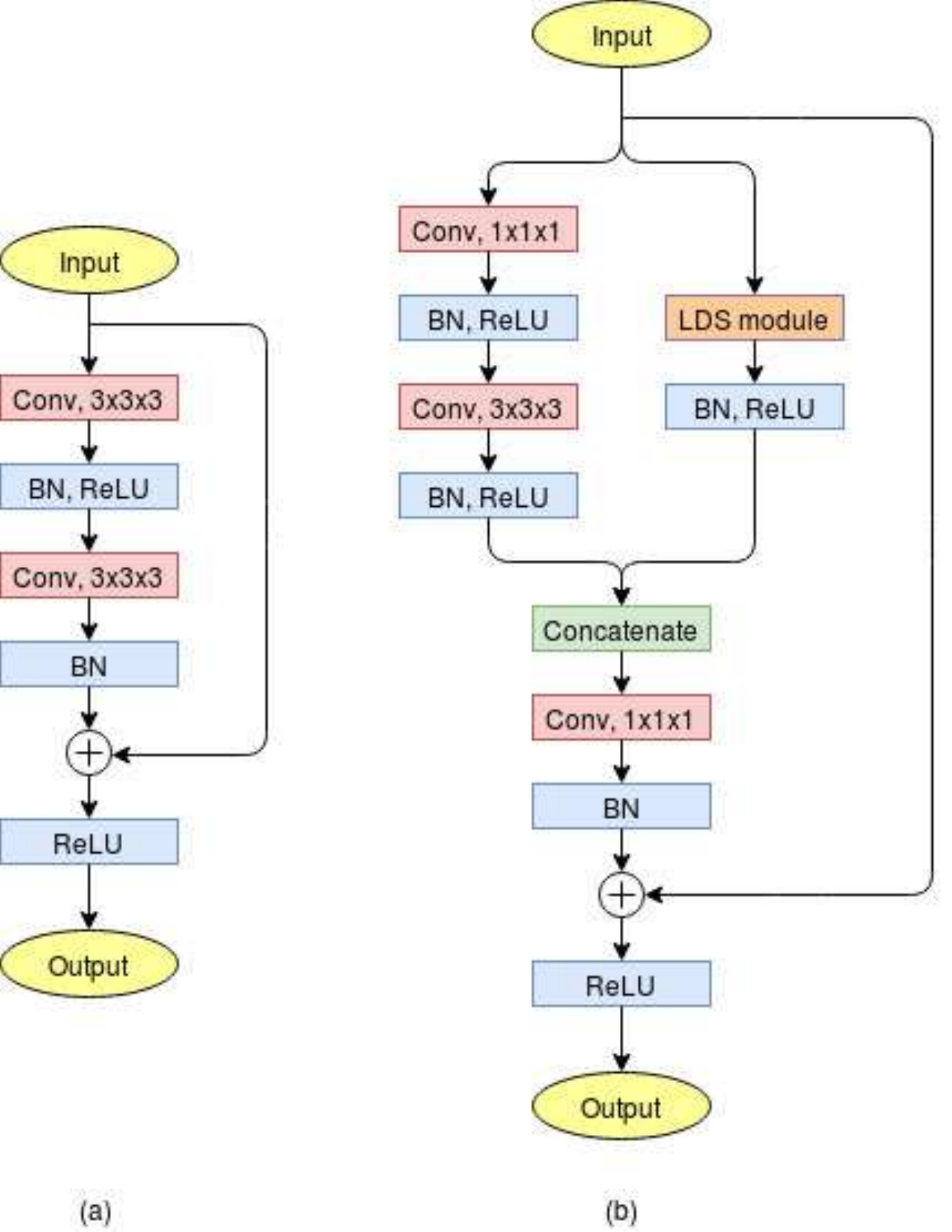}
    \caption[]{Baseline blocks for (a) original 3D-ResNet, (b) proposed bottleneck 3D-LDS-ResNet }\label{fig:resnet}
\end{figure}

As previously noted, the 3D-CNN in DeepSurf can be substituted by any 3D convolutional network of user's choice. In this work we adopted the ResNet \cite{he2016deep} architecture, which belongs to the family of residual networks. The main attribute of ResNet is the existence of skip connections between adjacent layers, so as to avoid the vanishing gradient problem. The baseline residual block of 3D-ResNet is depicted in Fig.~\ref{fig:resnet}(a). ResNet is formed by stacking a number of these blocks. We employed here a 18-layer ResNet, with the exact structure being shown in the original work \citep{he2016deep}.  
Considering the fact that we are employing 3D convolutions, the number of parameters in 3D-ResNet can be dramatically increased compared to 2D-ResNet. In the same work \cite{he2016deep}, the bottleneck architecture had also been presented, which allows more effective training of deeper ResNets with considerably less parameters per block. Recently, a novel residual network has been proposed, called LDS-ResNet \cite{dimou2018lds}, that has shown better performance than the baseline ResNet in computer vision tasks. Notably, LDS-ResNet acquired its best results when combined with a bottleneck architecture, which significantly surpassed all the non-bottleneck variants. In this work, we implemented a 3D variant of the bottleneck LDS-ResNet, with its main block depicted in Fig.~\ref{fig:resnet}(b). The difference to Fig.~\ref{fig:resnet}(a) is the addition of a second branch with an LDS module parallel to the original convolutional branch and the subsequent concatenation of these two branches. The extension of the LDS module in three dimensions, which is proposed in this paper, is illustrated in the following.

\subsubsection{Bottleneck 3D-LDS-Resnet}\label{sec:lds}

\begin{figure*}[!t]\centering
    \includegraphics[width=1\textwidth]{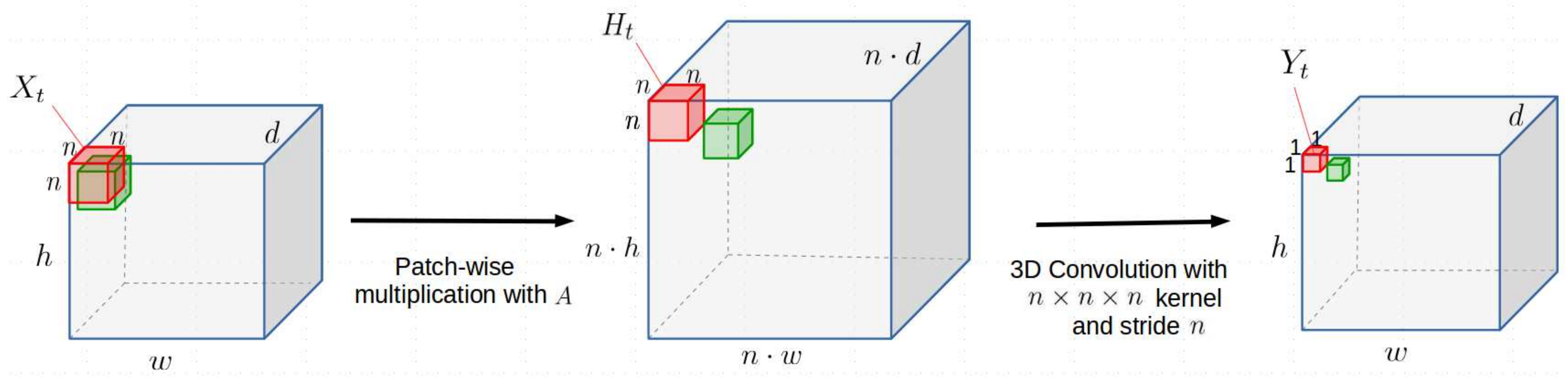}
    \caption[]{Graphical illustration of LDS-module's operation in three dimensions. For presentation reasons, only one input and output channel is considered ($d_{in}=d_{out}=1$).}\label{fig:lds}
\end{figure*}

LDS-ResNets were inspired by the Linear Dynamical Systems theory where a dynamical system is modeled through two time-evolving stochastic processes. The first process estimates a hidden state vector $h_t$ and the second one provides the observed output $y_t$ as a function of this hidden state. A similar approach was adopted in the LDS-module proposed by \cite{dimou2018lds}, with the exclusion of the time-evolution factor. The herein proposed 3D variant of this module is illustrated in Fig.~\ref{fig:lds}. Let us assume the input $X$ to the module is a 4D tensor of size $h\times w\times d\times d_{in}$, where \textit{h}, \textit{w}, \textit{d} are the spatial dimensions and $d_{in}$ is the number of channels. For clarification reasons, the operation is presented in Fig.~\ref{fig:lds} just for one channel ($d_{in}=1$). The LDS module operates iteratively over \textit{X} on 4D patches $X_t\in R^{n\times n\times n\times d_{in}}$ (in our experiments we used $n=3$). The calculation of the LDS module’s output $Y_t$ involves two main steps. 

The first one simulates the hidden state calculation of the LDS theory. Each patch $X_t$ is unfolded to a 2D matrix $x_t\in R^{n^3\times d_{in}}$ and the hidden state $h_t\in R^{n^3\times d_{in}}$ is obtained by:

\begin{equation}
h_t=Ax_t %\text{,}
\end{equation}
where $A\in R^{n^3\times n^3}$ is the hidden state transition matrix. Its values are randomly initialized for each layer and subsequently optimized during training through backpropagation. Then, $h_t$ is folded back to $H_t\in R^{n\times n\times n\times d_{in}}$ and for every \textit{t} these subvolumes are stored successively without overlapping, resulting in the intermediate volume of $H\in R^{nh\times nw\times nd\times d_{in}}$.  

The second step of the module performs the mapping from the hidden state $h_t$ to output $y_t$, as in original LDS theory. Specifically,

\begin{equation}
y_t=f\left(W,h_t\right)%\text{,}
\end{equation}
where \textit{f}() is a non-linear function with learnable parameters \textit{W}, and is implemented here by a convolutional operation. Volume \textit{H} is convolved with a set of $d_{out}$ filters $W\in R^{n\times n\times n\times d_{in}}$ with a stride $k\cdot n$ in each spatial dimension in order to align the filters \textit{W} with the regions corresponding to each of the patches $X_t$. Factor $k$ controls the downsampling rate of the specific building block. When $k=1$, the output $Y$ of the LDS module is a $h\times w\times d\times d_{out}$ tensor, while $k>1$ downsamples all spatial dimensions of the input tensor by a factor of $k$.

\section{Materials}\label{sec:materials}

The demonstrated efficiency of deep neural networks on many research fields lies greatly on the exploitation of large amount of qualitative and properly labeled data that can be used for training. The largest and most suitable database currently available for the BSP task is the scPDB database \cite{desaphy2014sc}, a continuously updated collection of ligandable binding sites of the Protein Data Bank. These binding sites are defined from complexes between a protein and a pharmacological ligand. One asset of scPDB is that, beyond the atom-based description of the protein and its ligand, it provides also their binding site, being thus suitable for a robust comparison and assessment of the examined methods. We utilized the 2017 release of the database which comprises 16034 entries corresponding to 4782 proteins with 17594 total binding samples. After removing some entries due to failure in reading or in feature extraction, the final dataset contains 15182 structures. For training and validation purposes, the remaining structures were split to 5 folds according to their Uniprot IDs, so as structures from the same protein should be included in the same fold. This separation ensures that the same protein pockets does not coexist in the training and testing set of a split, allowing a more robust and fair assessment.

For testing purposes, three different datasets were used, namely the COACH420, the HOLO4K and the apo/holo subsets from the CHEN dataset. COACH420 has been derived from the COACH test set \citep{yang2013protein} and consists of 420 single chain structures containing a mix of drug targets and naturally occurring ligands. HOLO4K is a larger dataset (4009 structures) containing larger multi-chain structures and was initially utilized by \cite{schmidtke2010large}. Finally, the CHEN dataset \citep{chen2011critical} was designed to cover a wide range of non-homologous protein structures and to include structures with the largest number of annotated binding sites.

\section{Implementation}\label{sec:implementation}

The solvent accessible surface (SAS) of proteins is calculated by the DMS software \citep{dms}. DMS returns a set of surface points along with the corresponding normal vectors. Despite setting the density parameter of DMS to a low value ($d=0.2$), the returned set of points is still quite dense, with the average minimum distance between neighboring points being 0.7 \angstrom. Parameter $f$, that controls the subsequent simplification process, should be set in a way that achieves a compromise between losing valuable surface information and avoiding excessive computational cost. In our case, we chose a value of $f=10$, which raised the average minimum distance of the remaining surface points to 2.3 \angstrom.

Prior to importing in DeepSurf, proteins should also be suitably pre-processed. Specifically, water, ions and ligands are removed from the PDB structures, and the remaining structure is protonated, if needed, enabling the proper computation of the necessary input features for the 3D-CNN~\citep{stepniewska2018development}. Before the final step of binding sites extraction (step 12 in Algorithm~\ref{alg:description}), hydrogen atoms are removed from the protein in order binding sites to maintain only heavy atoms. 

As previously stated, BSP is treated here as a binary classification problem, where the two considered classes are the "binding" and "non-binding" ones. Therefore, the 3D grids used as input data for training the 3D-CNN should belong to one of these classes. For each protein of scPDB, surface points that are within 4 \angstrom$ $ distance from any ligand atom are considered as binding points and the 3D grids localized on them are considered as samples of the "binding" class. Respectively, the 3D grids of the remaining surface points are considered samples of the "non-binding" class. In this case, the resulting dataset would be quite imbalanced, since the non-binding samples outnumber by far the binding ones.The class imbalance problem is a well-known problem in machine learning applications and a number of tactics have been proposed to tackle with it \citep{johnson2019survey}. The most common tactic lies on the data level and consists of either undersampling the main class or oversampling the secondary one. Due to the required time efficiency during training, the former technique was herein followed. For each protein, from the set of non-binding samples a number equal to the binding samples was randomly chosen in order to obtain a 50/50 balance between the two classes. 

DeepSurf was implemented in Python and the Tensorflow framework was employed for the deep learning operations. As already shown in Fig.~\ref{fig:lds}, the LDS module consists of two layers: a custom layer and a 3-D convolutional layer. The custom layer consists of the transition matrix $A$ calculation and the patch level multiplication (1). Since this is a patch-based iterative operation, like convolution, it can become extremely computationally heavy for larger input sizes. For this reason, the custom layer was implemented in CUDA to enable high level of parallelization and was, afterwards, integrated in Tensorflow. The source code of the method along with trained models are available at https://github.com/stemylonas/DeepSurf.git.

Regarding the training process, L2 regularization was applied on the weights of all convolutional layers ($\lambda=10^{-4}$), while batch normalization was applied with its default parameters. All models were trained for 20 epochs, with batch size of 64 samples, and were optimized by the Adam optimizer \citep{kingma2014adam} with a learning rate of $10^{-3}$.

\section{Results and Discussion}\label{sec:results}

The evaluation criteria used to assess the performance of the proposed method are the following:

\begin{itemize}
    \item \textbf{DCC:} Distance between the predicted and the real binding site center.
    \item \textbf{DCA:} Distance between the predicted binding site center and the closest ligand atom.
    \item \textbf{OVR:} Intersection on the atom level between the real and predicted binding sites divided by their union.
\end{itemize}

The DCC and DCA metrics have been widely used in previous works \citep{chen2011critical,jimenez2017deepsite,krivak2018p2rank}, to evaluate the localization quality of extracted binding sites by measuring their distance from either the annotated binding site or the corresponding ligand. A predicted binding site is considered as \textit{successful prediction} if the corresponding distance (DCC or DCA) is below a cutoff threshold $D_{cut}$. In all our experiments, we adopted a threshold of 4 \angstrom$ $. Finally, in order to assess the performance on the multi-protein level of a dataset, we provide in the following Tables success rates (\%), defined as the total number of \textit{successful predictions} for all proteins divided by the corresponding total number of existing sites.

On the other hand, OVR differs from the above distance-based metrics by considering also the shape of the binding sites, since it expresses a normalized spatial overlap between the predicted and the actual location of the binding pocket. In the following experiments, the DCC metric is used for evaluating the cross-validation performance on scPDB, since it is the only dataset with annotated binding sites, while the DCA and OVR metrics are employed for the comparative assessment on the three testing datasets. In all cases, the top-n and top-(n+2) predicted pockets are considered, where \textit{n} is the number of ligands for the specific protein. Finally, the ligandability threshold $T$ is set to 0.9 in all experiments. The sensitivity of DeepSurf on selection of $T$ is examined more thoroughly in Section~\ref{sec:T_sensitivity}.

\begin{table*}[!t]
\centering
\caption{Evaluation of DeepSurf using different network architectures. The number of parameters for each network is depicted along with the average cross-validation success rates (\%) on scPDB dataset using the DCC criterion ($D_{cut}=4$ \angstrom).}\label{Tab:scpdb} 
\vspace*{3mm}
{\begin{tabular}{lccccc}\hline 
 & $\#$params (M) && Top-n ($\%$) && Top-(n+2) ($\%$)\\\hline
Shallow network (DeepSite) & 1.0 && 62.1 && 64\\
ResNet-18 (w/o align) & 33.1 && 66.8 && 69.1\\
ResNet-18 (w align) & 33.1 && 68.1 && 70.4\\\hline
Bottleneck ResNet-18  & 1.7 && 66.8 && 69.4\\
Bottleneck ResNet-34  & 2.9 && 67.6 && 69.9\\
Bottleneck LDS-ResNet-18 & 3 && 68.3 && 70.8\\\hline
\end{tabular}}{}
\end{table*}

\subsection{Cross-validation on scPDB}\label{sec:cv}

The first stage of our experimentation consists of the 5-fold cross validation (CV) on scPDB. The goal of the conducted experiments is twofold. Firstly, we would like to evaluate the separate contribution of some fundamental steps of our method, such as the surface representation and the surface grid alignment, and, secondly, to test the behavior of DeepSurf with residual architectures of different size and type. As described in Section~\ref{sec:materials}, the scPDB dataset was split to five folds and for each fold a different model was trained. The obtained average performances on these folds and for all experiments are depicted in Table \ref{Tab:scpdb}. The corresponding number of network parameters is also included.

In the first experiment, we employed DeepSurf with the exact network architecture used in DeepSite \citep{jimenez2017deepsite} in order to demonstrate the added value of our proposed representation, while in the second one, we examined the contribution of the surface grid alignment step by training our method with (\textit{w align}) and without this feature (\textit{w/o align}). Although, in the original work of DeepSite, there is no exact reference to the obtained CV performance, we derived from the provided results that the average CV performance of DeepSite is about 50$\%$. Here, DeepSurf when using the same network architecture achieves a clearly better performance of 62.1$\%$ proving thus the effectiveness of the surface-based representation in comparison to the strict voxelization of the entire protein used in DeepSite. When the smaller network of DeepSite is replaced by the larger ResNet-18 architecture, we observe that DeepSurf achieves a higher top-n prediction score of 68.1$\%$. This finding indicates, as expected, that larger architectures are able to exploit more efficiently the large amount of training data and can provide much better generalization accuracies. When omitting the surface grid alignment step, the average performance drops by 1.3$\%$ in both top-n and top-(n+2) accuracies. This decrease is an indication that this feature, with no additive cost, can boost the overall performance of DeepSurf.

We also conducted additional experiments with alternative lightweight residual architectures in place of ResNet-18 in order to investigate whether lighter networks can achieve similar performances. As already stated in Section~\ref{sec:implementation}, the number of parameters in 3D-ResNets can be quite large due to the 3D convolutions, e.g. ResNet-18 has about 33 million parameters. When replacing the basic residual block with the respective bottleneck (Fig.~\ref{fig:resnet}), the resulted network has significantly less parameters (1.7 million), yet it leads to an expected drop in accuracy of 1-1.3$\%$. The addition of LDS block (bottleneck LDS-ResNet-18), which was detailed in Section~\ref{sec:lds}, led to a rise of 1.5$\%$ in accuracy comparing to bottleneck ResNet-18, but at the cost of approximately double parameters. For a fairer assessment of the contribution of the LDS block, we tested DeepSurf employed with a bottleneck ResNet with 34 layers, which has the same parameters as the bottleneck LDS-ResNet-18. We notice that the LDS variant achieved 0.7$\%$ higher top-n accuracy and 0.9$\%$ higher top-(n+2) accuracy than bottleneck ResNet-34, making it preferable as a lightweight architecture. Comparing now to the baseline ResNet-18 variant, although bottleneck LDS-ResNet-18 has more than 10 times fewer parameters, it achieves similar, if not better, CV performance than its competitor. 

\begin{table*}[!t]
\caption{Performance comparison of DeepSurf and the competing DL methods using the DCA criterion ($D_{cut}=4$ \angstrom).}
\label{Tab:compare_dl} 
\centering
\vspace*{3mm}
{\begin{tabular}
{@{}lcccccc@{}}\hline 
 && \multicolumn{2}{c}{COACH420} && \multicolumn{2}{c}{HOLO4K}\\\cline{3-4}\cline{6-7}
 && Top-n & Top-(n+2) && Top-n & Top-(n+2)\\\hline
DeepSite \cite{jimenez2017deepsite}  && 57.5 & 65.1 && 45.6 & 48.2\\
Jiang \textit{et al.} \cite{jiang2019novel} && 55 & 58.7 && 38.2 & 41.5\\
Kalasanty \cite{stepniewska2020improving}  && 68 & 70.4 && 32.1 & 32.3\\
DeepSurf (ResNet-18)  && 72.1 & 73.3 && 50.1 & 50.6\\
DeepSurf (Bot-LDS-ResNet-18) && 71.7 & 72.7 && 50.4 & 50.8\\\hline
\end{tabular}}{}
\end{table*}

\begin{table}[!t]
\caption{Performance comparison of DeepSurf and Kalasanty using the OVR criterion, computed only for correctly located binding sites ($DCA < 4$ \angstrom).}\label{Tab:ovr} 
\vspace*{3mm}
{\begin{tabular}
{@{}lcc@{}}\hline 
 & \small{COACH420} & \small{HOLO4K}\\\hline
Kalasanty  & 0.21 & 0.15\\
DeepSurf ({\footnotesize ResNet-18})  & 0.29 & 0.17\\
DeepSurf ({\footnotesize Bot-LDS-Res-18}) & 0.28 & 0.17\\\hline
\end{tabular}}{}
\end{table}

\begin{table*}[!t]
\caption{Qualitative comparison of DeepSurf and the competing DL methods. The number of proteins where each method failed to produce a pocket and the average number of predicted pockets are shown.}\label{Tab:qualitative} 
\centering
\vspace*{3mm}
{\begin{tabular}
{lcccccc}\hline 
 && \multicolumn{2}{c}{\parbox{10em}{\centering Number of \\ failures}} 
 && \multicolumn{2}{c}{\parbox{12em}{\centering Average number of \\ predicted pockets}}\\\cline{3-4}\cline{6-7}
 && COACH420 & HOLO4K && COACH420 (1.2) & HOLO4K (2.8)\\\hline
DeepSite \cite{jimenez2017deepsite}  && 3 & 21 && 3.2 & 2.8\\
Jiang \textit{et al.} \cite{jiang2019novel}  && 12 & 65 && 1.4 & 3.4\\
Kalasanty \cite{stepniewska2020improving}  && 16 & 475 && 1.1 & 1.2\\
DeepSurf (ResNet-18)  && 7 & 5 && 1.1 & 1.8\\
DeepSurf (Bot-LDS-ResNet-18) && 8 & 10 && 1.1 & 1.8\\\hline
\end{tabular}}{}
\end{table*}

\subsection{Comparison to DL-based methods}

After evaluating the individual features of our method through cross-validation, we perform comparison of DeepSurf to other competing in the BSP task deep learning methods that are publicly available. Specifically, we perform comparison to DeepSite \citep{jimenez2017deepsite}, Jiang \textit{et al.} \cite{jiang2019novel} and Kalasanty \citep{stepniewska2020improving}. From the various architectures of DeepSurf tested in Section~\ref{sec:cv}, we keep for comparison the baseline ResNet-18 and the lightweight bottleneck LDS-ResNet-18, which provided the highest accuracies. For testing purposes, the COACH420 and HOLO4K datasets were utilized (for more details see Section~\ref{sec:materials}). In order to avoid data leakage, a global sequence alignment between all targets from the training and the three test sets was performed, and any training target with more than 90\% sequence similarity with a testing one was removed. The remaining dataset, consisting of 9444 targets, was used to train the two variants of DeepSurf. Although all of the competing methods have been trained on the same database (scPDB), any proteins common to our testing datasets have not been removed. This means that these methods have a slight advantage due to this specific data leakage. In case that a method fails to produce any binding site, an adequately large value of DCA is assigned for each ligand of this protein ensuring that this solution will be regarded erroneous. 

The obtained DCA performances for the COACH420 and HOLO4K datasets are shown in Table~\ref{Tab:compare_dl}. The provided DeepSite results are those obtained by \cite{krivak2018p2rank}. Regarding the three competing methods, we notice that Kalasanty surpasses clearly the others in COACH420, while DeepSite is by far superior in the most challenging dataset of HOLO4K. Nevertheless, DeepSurf clearly outperforms all competing methods in both datasets. Specifically, DeepSurf is superior to Kalasanty in COACH420 by 4$\%$ in top-n accuracy and 3$\%$ in top-(n+2), while in HOLO4K, DeepSurf outperforms DeepSite by 4.5$\%$ in top-n accuracy and 2.5$\%$ in top-(n+2). In order to more thoroughly examine the generalization capabilities of DeepSurf, the test sets have been split into bins of various similarity ranges, based on their maximum global sequence similarity to the training set, and the obtained performances for each bin are depicted in Supplementary Figure S1. We can notice from both datasets that the performance of DeepSurf is slightly increased with increasing similarity of the test proteins. However, the generalizability of the method is demonstrated by considering the first bin, which corresponds to the most dissimilar structures (similarity less than 40\%). In this case, DeepSurf achieves 67\% in COACH420 and 44.5\% in HOLO4K, which are the highest results among the competing methods and about 5\% lower to its overall performance. From the above results, we can conclude that the two DeepSurf alternatives behave similarly when applied to unknown structures. This indicates the computational and generalization effectiveness of the LDS-equipped network, since it achieves similar results to ResNet-18 but with the benefit of more than 10 times fewer parameters.

For a more comprehensive comparison of the above methods, an overlapping criterion should also be applied that evaluates the shape of the extracted pockets. According to \cite{luscombe2001amino} and \cite{chen2011critical}, binding sites are defined as the  non-hydrogen atoms of a residue that are within 4 \angstrom$ $ to a non-hydrogen atom of the ligand. Following this principle, we extracted binding sites for all proteins in COACH420 and HOLO4K and computed the OVR values only for the correctly located binding pockets ($DCA < 4$ \angstrom) in each corresponding case. The obtained average values are presented in Table~\ref{Tab:ovr}. Except from DeepSurf, Table~\ref{Tab:ovr} holds also the average OVR values obtained by Kalasanty, since it is the only competing method that, additionally to centers, returns explicitly the binding site atoms. As we can see, DeepSurf achieves higher overlapping values in both datasets, and especially in COACH420. Nevertheless, the attained values, mainly in HOLO4k, are relative small compared to the ideal score of 1, indicating that the extraction of properly shaped binding sites is still an open issue.

A more qualitative assessment of the competing methods is given in Table~\ref{Tab:qualitative}, which provides the average number of predicted pockets per protein along with the number of proteins where each method failed to produce a single pocket. As we can see, Kalasanty was unable to extract binding sites for a large number of proteins, even after adjusting its default parameters. For example, in the case of HOLO4K, no binding site returned for 475 out of 4009 proteins. Among the two DeepSurf variants, ResNet-18 appeared more robust, since it encountered the fewer failures in case of HOLO4K. Regarding the number of extracted binding sites, DeepSurf and Kalasanty have the tendency to return fewer pockets than DeepSite and Jiang's method in both datasets. In the case of COACH420, it is beneficial since both methods extract a number of pockets similar to the average number of true ones (1.2). On the other hand, in HOLO4K, DeepSurf, and especially Kalasanty, return on average fewer binding sites than the actual ones (2.8). This can explain the larger differences between top-n and top-(n+2) accuracies observed in the case of DeepSite and Jiang's method compared to the rest of the methods (see Table~\ref{Tab:compare_dl}).    

\subsection{Comparison between apo and holo structures}

\begin{table*}[!t]
\caption{Performance comparison between apo and holo structures for DeepSurf and the competing DL-based methods using the DCA criterion ($D_{cut}=4$ \angstrom).\label{Tab:apo_holo}}
\centering
\vspace*{3mm}
{\begin{tabular}
{@{}lcccccc@{}}\hline 
 && \multicolumn{2}{c}{CHEN\_holo} && \multicolumn{2}{c}{CHEN\_apo}\\\cline{3-4}\cline{6-7}
 && Top-n & Top-(n+2) && Top-n & Top-(n+2)\\\hline
Jiang \textit{et al.} \cite{jiang2019novel} && 34.5 & 35 && 28.4 & 30.5\\
Kalasanty \cite{stepniewska2020improving}  && 35.5 & 36 && 33 & 34\\
DeepSurf (ResNet-18)  && 40.6 & 40.6 && 39.6 & 39.6\\
DeepSurf (Bot-LDS-ResNet-18) && 39.1 & 39.1 && 37.6 & 37.6\\\hline
\end{tabular}}{}
\end{table*}

Both previous datasets consist of holo structures, i.e., structures that are bound to ligands. However, because in real life scenarios the protein-ligand interactions may lead to conformational changes at the proximity of the binding site, we also investigate the performance of DeepSurf in case of apo structures, i.e., unbound-state proteins. For this reason, the apo and holo subsets of the CHEN dataset were used, which constitute of 104 holo proteins and their corresponding 104 apo structures \cite{chen2011critical}. For a proper evaluation on apo structures, they were initially structurally aligned on their holo complements and afterwards the corresponding holo ligands were assigned to them. The obtained DCA performances from both DeepSurf and the competing methods are depicted overall in Table~\ref{Tab:apo_holo}, and for various global sequence similarity ranges in Supplementary Fig. S2. We can see that across both subsets, the two DeepSurf variants achieve superior results than the competing methods. Especially, when considering the bin with the most dissimilar proteins, our proposed method achieves the highest performances among the competitors with accuracies between 35 and 38\%. Interestingly, someone can observe that, although all methods provide better results with increasing similarity, there is a significant drop in performance for DeepSurf and Kalasanty on the last bin with the most similar proteins. However, since the number of proteins in this bin is quite small (only 4), it prevents us from drawing a solid conclusion for the statistical significance of this result. When comparing the performances between holo and apo structures, we observe that, although all methods perform worse on the apo case, DeepSurf exhibits the smallest decrease in accuracy (1-1.5\%). Especially, in the case of the most dissimilar subset in Fig. S2, we notice that the superiority of DeepSurf from the second best competitor is even larger on the most challenging apo structures.

\begin{figure}[!t]\centering
\begin{tabular}{cc}
    \includegraphics[width=0.233\textwidth]{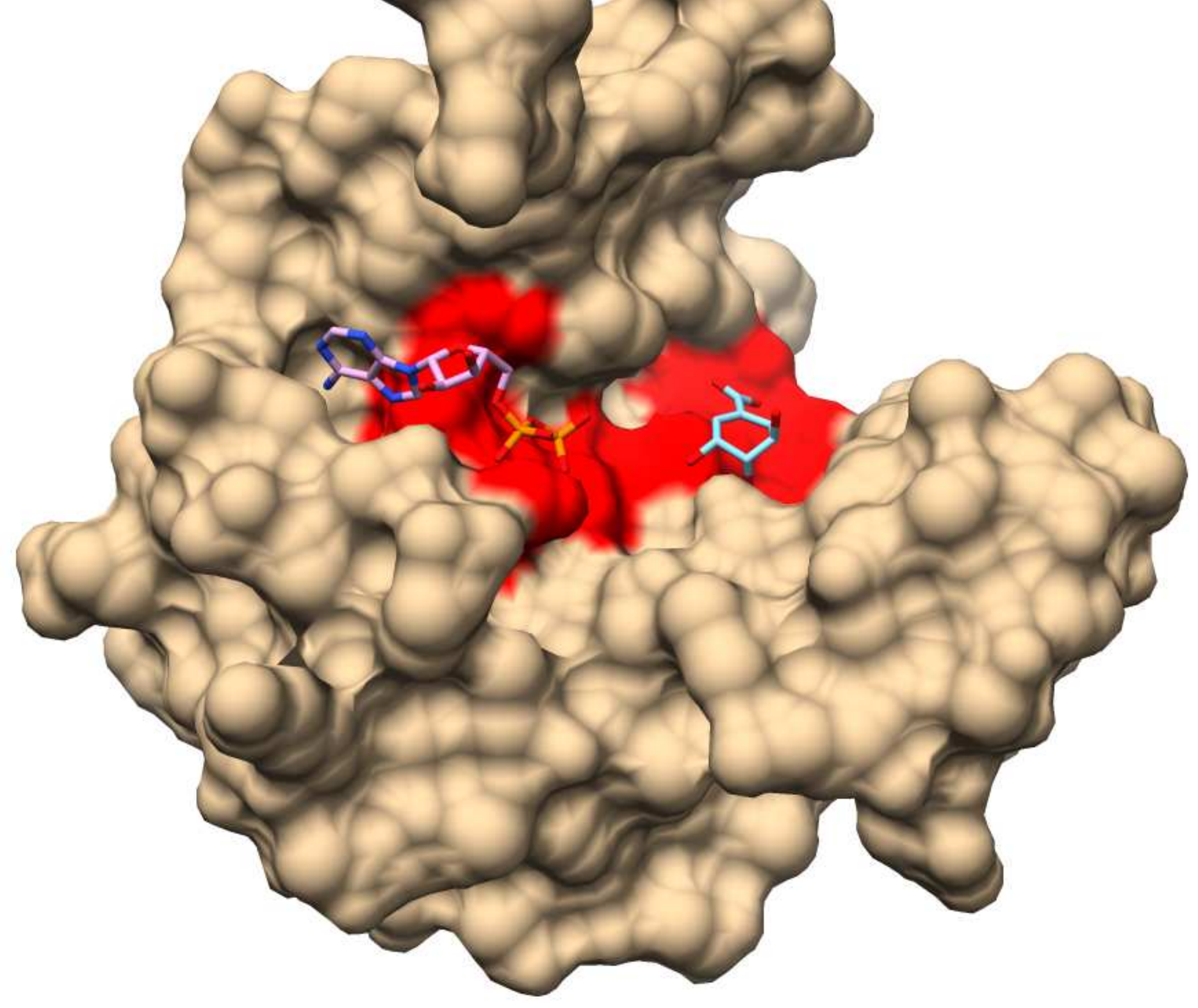} &   \includegraphics[width=0.215\textwidth]{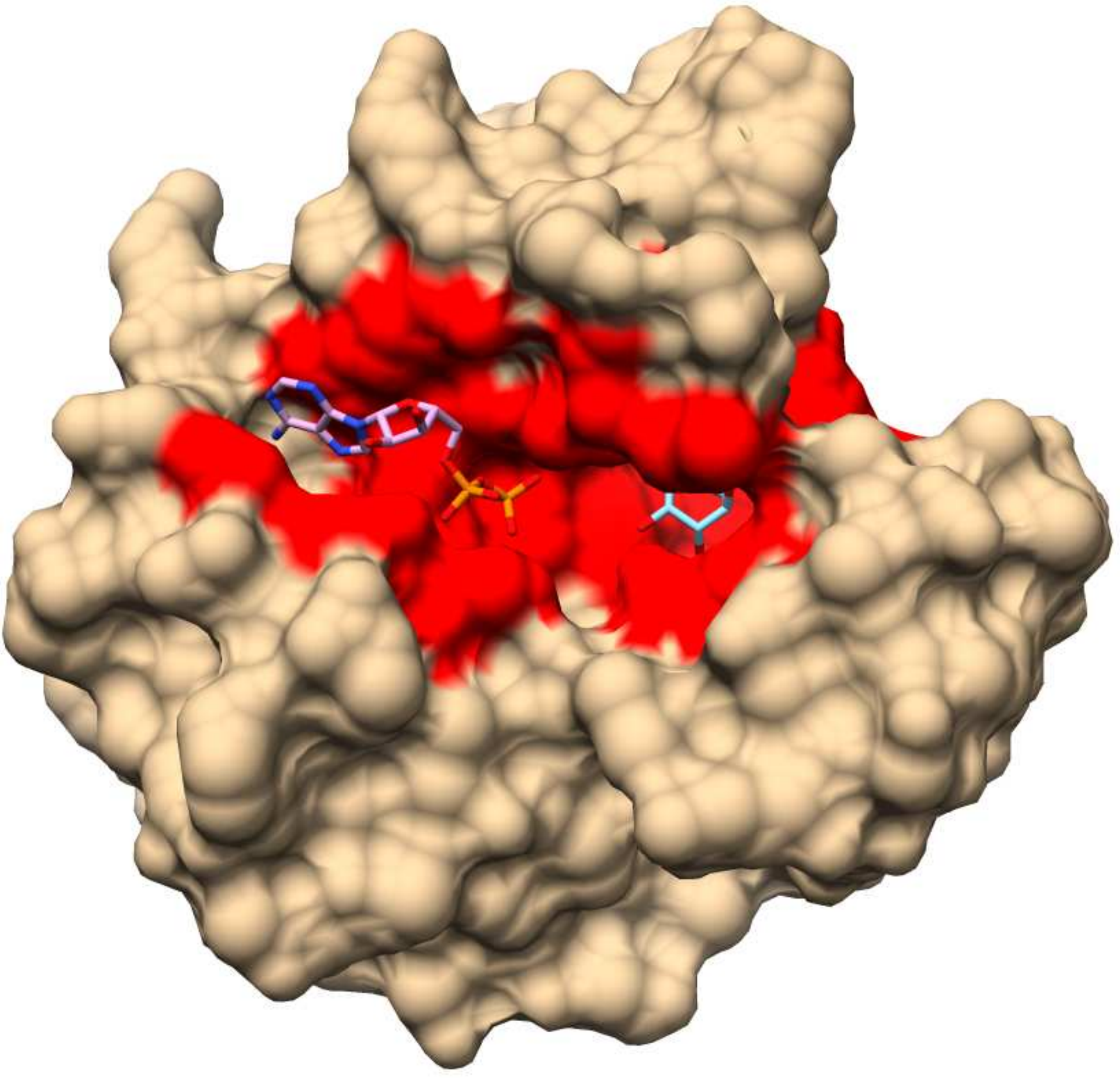} \\
    \quad (a) & \quad (b)
\end{tabular}
\caption[]{Extraction example of a cryptic site between (a) the apo structure of '2iyt' and (b) the holo structure of '2iyq'.}\label{fig:apo_holo}
\end{figure}

Fig.~\ref{fig:apo_holo} demonstrates the behavior of DeepSurf in the case of a cryptic binding site between the apo structure of '2iyt' and the holo structure of '2iyq'. Cimermancic \textit{et al.} \cite{cimermancic2016cryptosite} defined cryptic site as a site that forms a pocket in a ligand-bound structure but not in the corresponding unbound, and recognized a set of protein pairs that have cryptic binding sites. The herein chosen pair of proteins is included in this set. This can also be noticed visually in Fig.~\ref{fig:apo_holo}(b), where the protein in its holo form modifies its structure by folding down in order to enclose the bound ligands. Our proposed method detected the true binding site in both cases, and especially in the most challenging case of the '2iyt' where the binding pocket is not so well-formed as in the case of '2iyq'. Summarizing, although the obtained accuracies are quite small and there is still plenty room for improvement, the small reduction performed by DeepSurf on apo structures and the efficient detection of the cryptic site indicate the ability of the method to learn generic geometrical and physicochemical features, enabling it to bypass possible small local conformations on the proteins geometry.

\subsection{Sensitivity on ligandability threshold}\label{sec:T_sensitivity}

\begin{figure}[!t]
\begin{tabular}{cc}
    \includegraphics[width=0.22\textwidth]{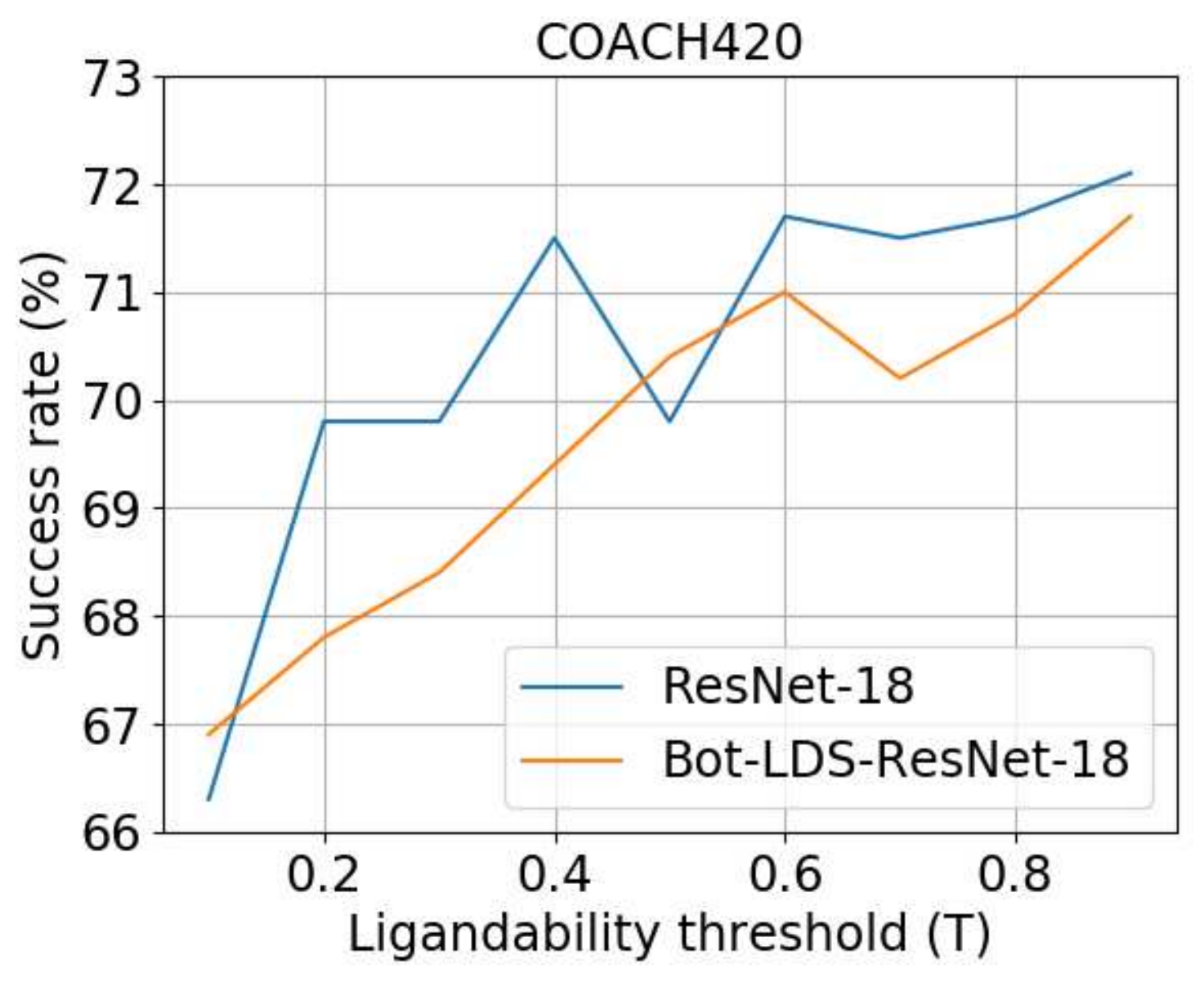} &   \includegraphics[width=0.22\textwidth]{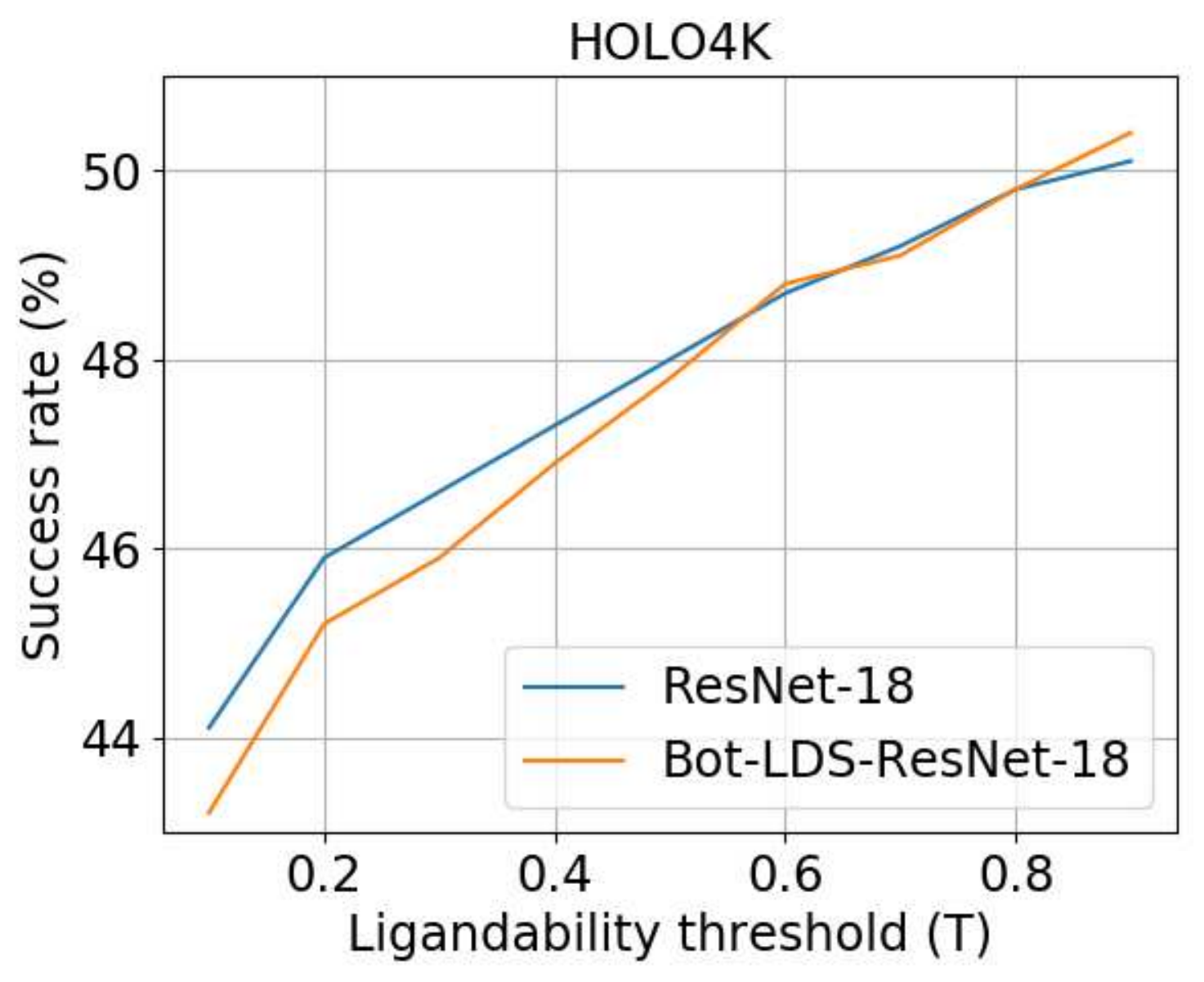} \\
    \quad (a) & \quad (b) \\[8pt]
    \includegraphics[width=0.22\textwidth]{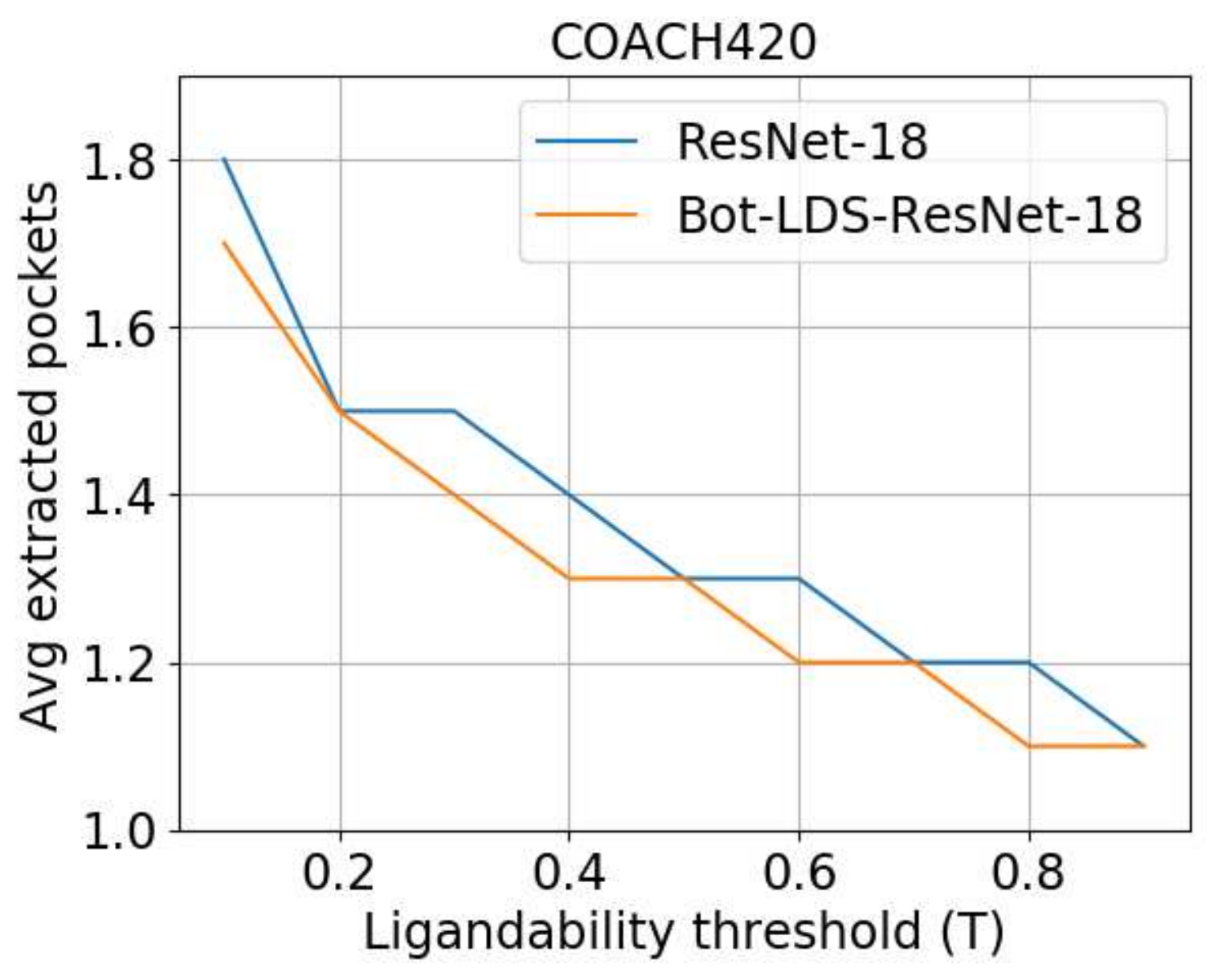} &   \includegraphics[width=0.22\textwidth]{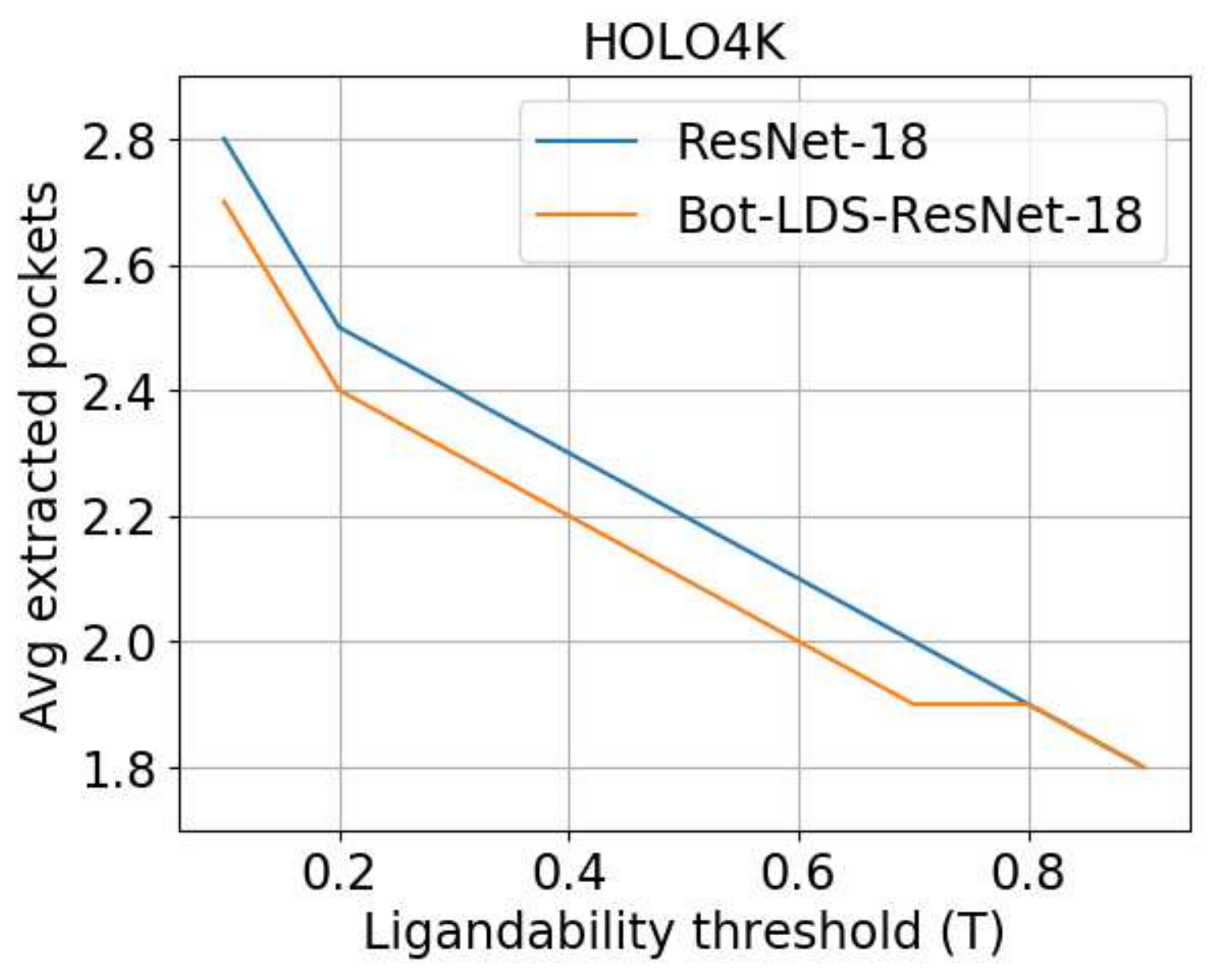} \\
    \quad (c) & \quad (d)
\end{tabular}
%\caption[]{Influence of varying $T$ (a), (b) on the obtained performance and (c), (d) on the average number of extracted pockets.}\label{fig:T}
\caption[]{(a), (b) Obtained performance and (c), (d) average number of extracted pockets for different ligandability thresholds.}\label{fig:T}
\end{figure}

A key parameter of DeepSurf is the ligandability threshold $T$ above which surface points are considered potential binders. Its influence on the obtained results is examined quantitatively in Fig.~\ref{fig:T}, where the success rates and the average number of extracted pockets for both DeepSurf variants and for various ligandability thresholds are presented. As we can see, lower values of $T$ lead to a general decline in performance for both datasets, and especially in the case of HOLO4K, where a consistent drop is noticed. On the other hand, decrease of $T$ leads to an expected raise in the number of extracted pockets, since more points are preserved across the protein surface.

In order to evaluate the significance of the previously observed differences between the various values of $T$, the statistical Wilcoxon signed-rank test was employed. Specifically, we performed pairwise tests between all possible pairs of $T$, by considering the DCA values obtained by DeepSurf in each case. DCA values greater than 10 \angstrom $ $ were assumed equally wrong and were hence truncated to 10 \angstrom. The tests were computed on the COACH420 dataset for both DeepSurf variants and the obtained results are presented in Table~\ref{Tab:stat} for $T>=0.4$. Lower values were omitted, since, as it was expected from Fig.~\ref{fig:T}, they were all proved significantly worse than higher $T$ values. The ‘+’ / ‘-’ indicates that a $T$ value in a given column is significantly better/worse than the $T$ value in the given row with $p < 0.05$, while ‘=’ denotes that a given pair of values are not significantly different. Among the two DeepSurf variants, we observe that the lighter one is more affected by variations of $T$, since $T=0.9$ provide significantly better results than lower values, whereas, in the case of ResNet-18, all threshold values larger than 0.6 are statistically equivalent. 

\begin{figure}[!t]\centering
\begin{tabular}{cc}
    \includegraphics[width=0.225\textwidth]{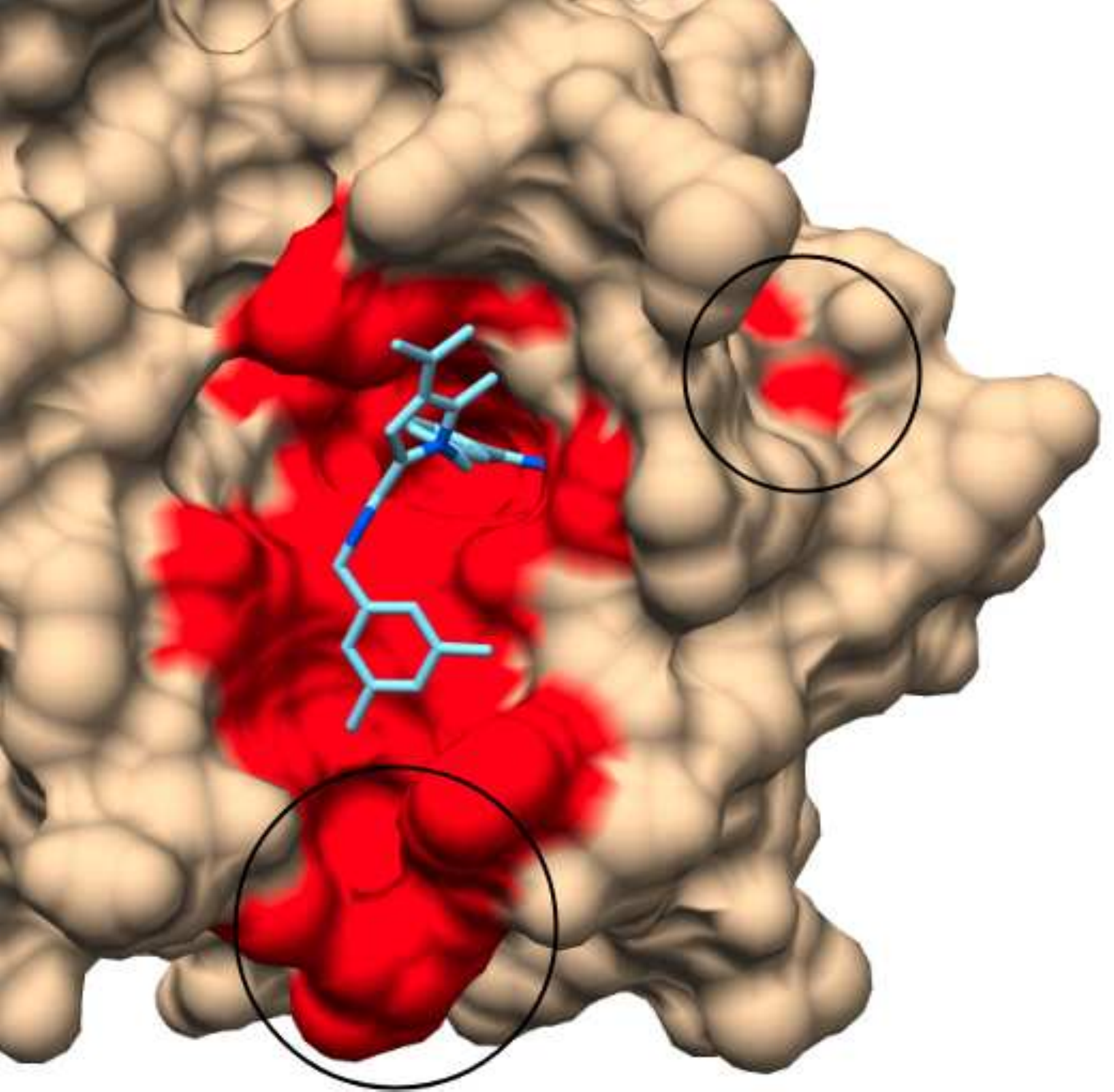} &   \includegraphics[width=0.225\textwidth]{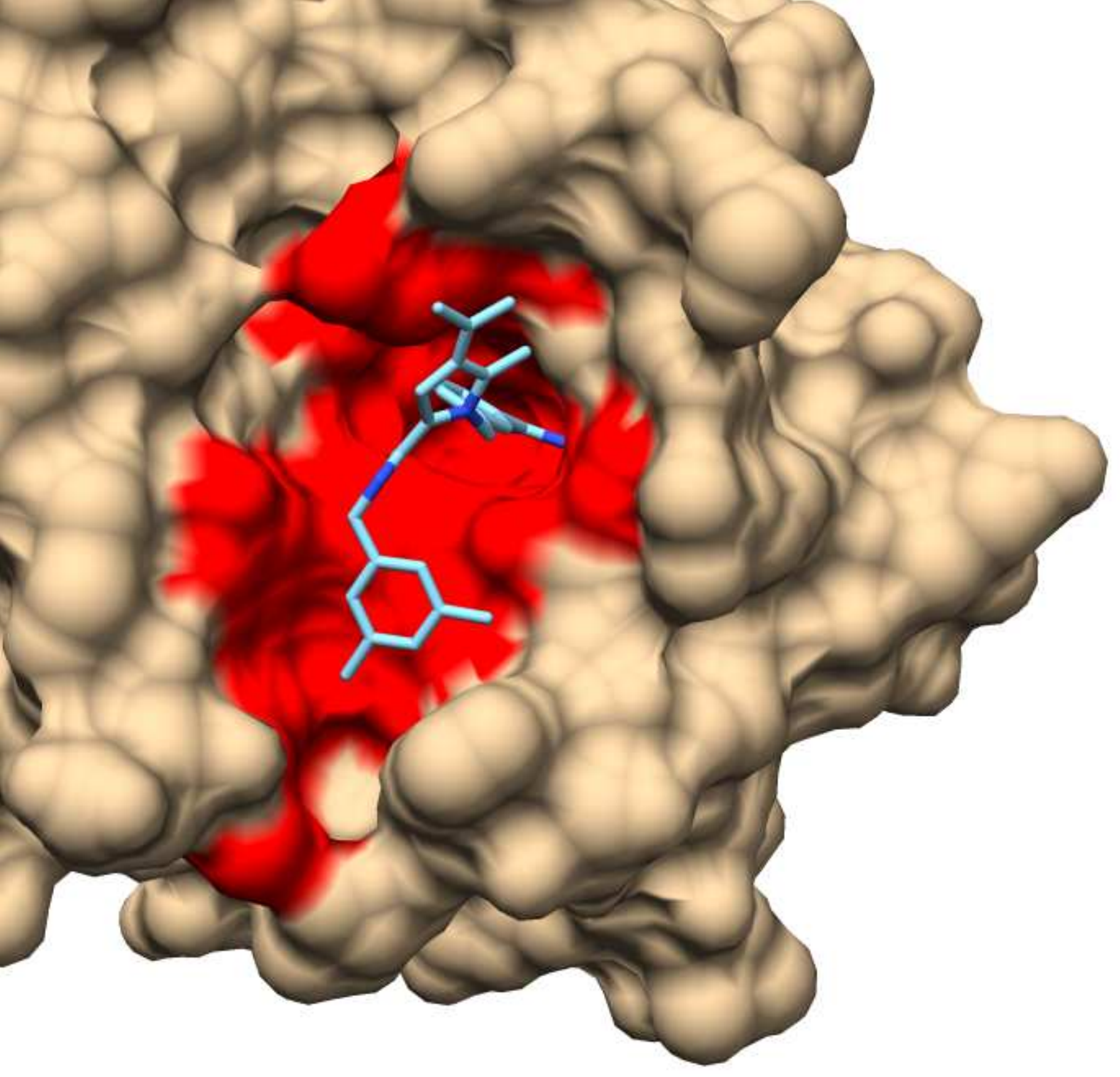} \\
    \quad (a) & \quad (b)
\end{tabular}
\caption[]{Binding site extraction example for structure '1lqdB' with ligandability threshold (a) $T=0.5$ and (b) $T=0.9$. Black circles point the areas where the two results differ.}\label{fig:bsite_T}
\end{figure}

In our previous experiments, a high ligandability threshold of $T=0.9$ was selected for both variants. This choice was not based only on the previous statistical evidence, but also on the visual inspection of the extracted pockets. An illustrative example is given in Fig.~\ref{fig:bsite_T}, which displays the binding sites extracted by DeepSurf for structure '1lqdB' with $T=0.5$ and $T=0.9$, respectively. Although, in both cases, the extracted pockets are considered successful due to low DCA values (3.6 and 3.2 \angstrom$ $ respectively), we can observe that the extracted pocket in Fig.~\ref{fig:bsite_T}(a) is larger and expands to undesired areas (marked with black circles) away from the ligand. This is totally expected, since a smaller value of $T$ leads to the preservation of more surface points before clustering and, subsequently, to the formation of larger binding sites. From the aforementioned, it is concluded that DeepSurf exhibits its optimal performance when high ligandability thresholds are set. However, based on the previous statistical study, when ResNet-18 variant is employed, $T$ can be lowered down to 0.6 in cases where larger binding sites are required.

\begin{table*}[!t]
\caption{Statistical significance of the differences in DCA between various ligandability thresholds using Wilcoxon signed-rank test ($p=0.05$).\label{Tab:stat}} 
\centering\vspace*{3mm}
{\begin{tabular}
{lccccccccccccccccc}\hline 
&&& \multicolumn{6}{c}{ResNet-18} &&&& \multicolumn{6}{c}{Bot-LDS-ResNet-18}
\\\cline{4-9}\cline{13-18}
$T$ &&& 0.4 & 0.5 & 0.6 & 0.7 & 0.8 & 0.9 &&&& 0.4 & 0.5 & 0.6 & 0.7 & 0.8 & 0.9\\\hline
0.4 &&& & = & + & + & + & + &&&& & = & + & + & + & +\\
0.5 &&& = & & = & = & = & = &&&& = & & + & + & + & +\\
0.6 &&& - & = & & = & = & = &&&& - & - & & = & = & +\\
0.7 &&& - & = & = & & = & = &&&& - & - & = & & = & +\\
0.8 &&& - & = & = & = & & = &&&& - & - & = & = & & +\\
0.9 &&& - & = & = & = & = & &&&& - & - & - & - & - &\\\hline
\end{tabular}}{}
\end{table*}

\subsection{Comparison to non-data-driven methods}

As a final assessment of DeepSurf, we compared its performance to a set of traditional non-data-driven approaches, such as the geometry-based Fpocket \citep{capra2009predicting}, the energy-based AutoSite \citep{ravindranath2016autosite} and the template-based COFACTOR \citep{roy2012cofactor}. The obtained success rates are depicted in Table~\ref{Tab:compare_nondl}. COFACTOR, as a template-based method, searches in a large database of known protein-ligand complexes to identify similar binding sites to a query protein. Previous to its application, we applied the same filtering procedure as in our training set in order to accomplish a fairer comparison to the rest of the methods. Specifically, before each query search, identical and close homologues of more than 90\% sequence similarity to the query were excluded from the database. Due to high execution times, COFACTOR was tested only on the smaller COACH420 set. The other two methods were applied with their default parameters. We performed also a generalizability analysis, where the test sets were partitioned into homologous and non-homologous subsets using a threshold of 40\% global sequence similarity, and the corresponding performances are shown in Supplementary Fig. S3.

\begin{table*}[!t]
\caption{Performance comparison of DeepSurf with competing non-data-driven methods using the DCA criterion ($D_{cut}=4$ \angstrom).\label{Tab:compare_nondl}} 
\centering\vspace*{3mm}
{\begin{tabular}
{lcccccc}\hline 
 && \multicolumn{2}{c}{COACH420} && \multicolumn{2}{c}{HOLO4K}\\\cline{3-4}\cline{6-7}
 && Top-n & Top-(n+2) && Top-n & Top-(n+2)\\\hline
Fpocket \citep{capra2009predicting}  && 41.6 & 56.6 && 40.3 & 47.1\\
AutoSite \citep{ravindranath2016autosite} && 56.4 & 69.9 && 51.8 & 58.3\\
COFACTOR \citep{roy2012cofactor} && 70.7 & 78.7 && - & -\\
%P2Rank   && 71.9 & 78.5 && 56.9 & 61.3\\
DeepSurf (ResNet-18)  && 72.1 & 73.3 && 50.1 & 50.6\\\hline
%DeepSurf (Bot-LDS-ResNet-18) && 71.7 & 72.7 && 50.4 & 50.8\\\hline
\end{tabular}}{}
\end{table*}

\begin{table*}[!t]
\caption{Qualitative comparison of DeepSurf and the competing non-data-driven methods. The number of proteins where each method failed to produce a pocket and the average number of predicted pockets are shown.\label{Tab:qualitative_nondl}} 
\centering\vspace*{3mm}
{\begin{tabular}
{lccccc}\hline 
 & \multicolumn{2}{c}{\parbox{6em}{\centering Number of \\ failures}} 
 && \multicolumn{2}{c}{\parbox{10em}{\centering Avg number of \\ predicted pockets}}\\\cline{2-3}\cline{5-6}
 & \small{COACH420} & \small{HOLO4K} && \small{COACH420 (1.2)} & \small{HOLO4K (2.8)}\\\hline
Fpocket \citep{capra2009predicting}  & 0 & 1 && 13.9 & 23.8\\
AutoSite \citep{ravindranath2016autosite}  & 1 & 4 && 13.6 & 24\\
COFACTOR \citep{roy2012cofactor}  & 2 & - && 7.3 & -\\
DeepSurf (ResNet-18)  & 7 & 5 && 1.1 & 1.8\\\hline
\end{tabular}}{}
\end{table*}

We initially observe that Fpocket is clearly inferior to the rest of the methods in both datasets. Regarding AutoSite, although it achieves much lower accuracies than COFACTOR and DeepSurf in COACH420, it performs considerably better on HOLO4k by achieving the highest accuracies, and especially when top-(n+2) solutions are examined. On the other hand, COFACTOR performs slightly worse than DeepSurf in top-n accuracy, while it is clearly superior when more solutions are considered. However, it should be noted that COFACTOR, as all template-based methods, suffers from high execution times (order of hours) instead of the much faster DeepSurf (order of seconds). Similar findings to Table~\ref{Tab:compare_nondl} can be noticed when regarding the non-homologous subsets in Fig. S3. Although DeepSurf achieves competitive results, it is edged in both cases by a non-data-driven method. Finally, we can notice that, besides Fpocket, the rest of the methods are favored when tested on the most similar proteins of the homologous subset, maintaining in great extent the relative difference observed in the non-homologous subset. This can imply that the better performance of DeepSurf between the two subsets may be due to intrinsic characteristics of the protein sets and not the underlying similarities, since even non-data-driven methods, as AutoSite, exhibit similar increase in performance.

One major observation is that all competing methods obtain considerably higher top-(n+2) accuracies comparing to the corresponding top-n ones, in contrast to DeepSurf which exhibits only a slight increase of 0.5 to 1\%. This difference can be attributed to the large number of extracted pockets by these methods as shown in Table~\ref{Tab:qualitative_nondl}. For example, in the case of COACH420, Fpocket and AutoSite extracts approximately 14 pockets while COFACTOR extracts 7. On the contrary, as shown in Table~\ref{Tab:qualitative}, all DL-based methods, including DeepSurf, output a small number of binding sites, closer to the actual one. This tendency is probably induced by restrictions in the training set, such as the incompleteness of the training data in combination to inherent peculiarities of the BSP task. For example, a surface grid currently labeled as negative sample due to the absence of known binding ligand on this site, could be a positive sample in practice due to an unknown so far binding. Continuous enhancement of the training database with more pocket samples from more diverse binding cases could lead to performance improvement.

\begin{figure*}[!t]\centering
\begin{tabular}{cccc}
    \includegraphics[width=0.238\textwidth]{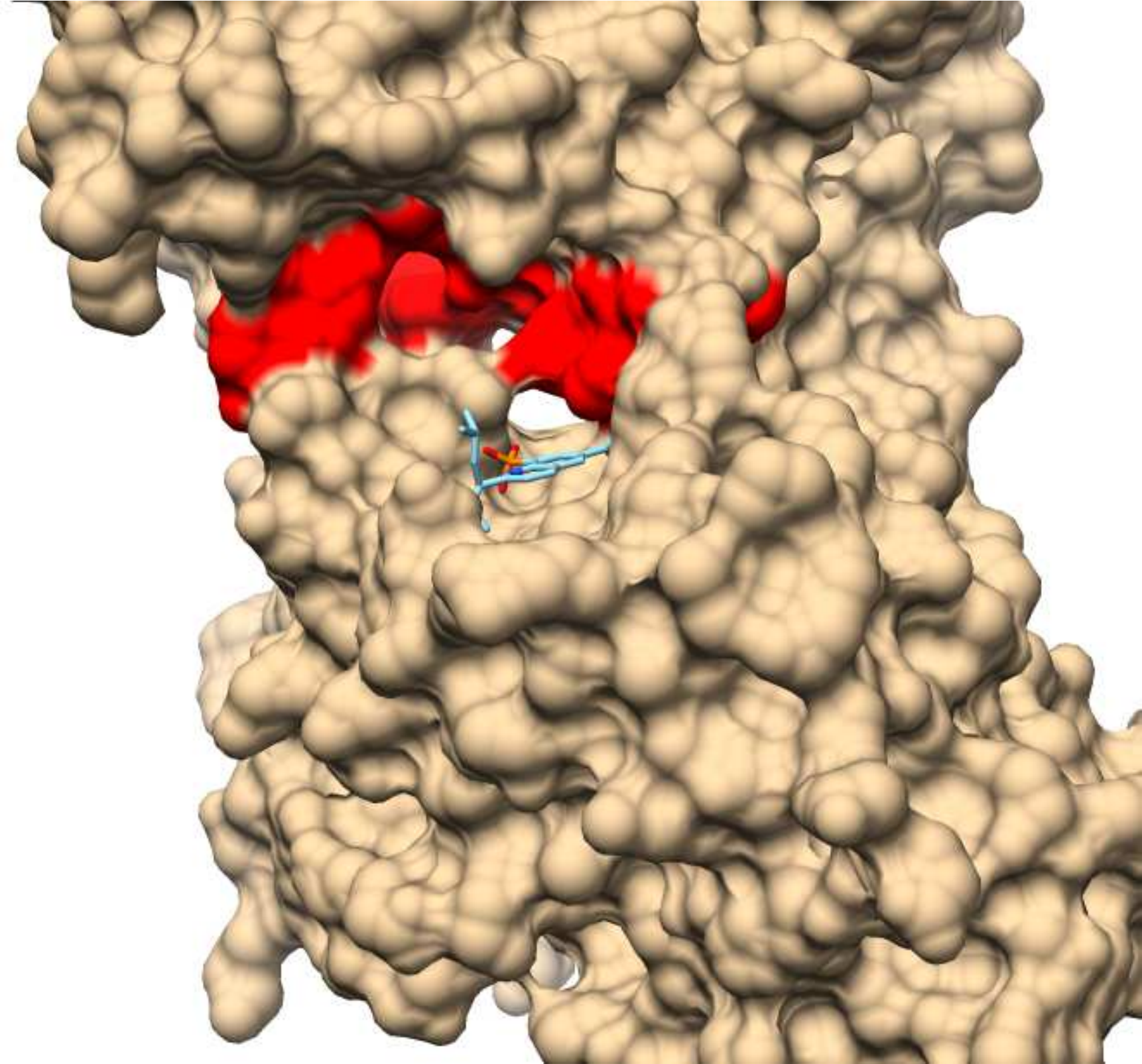} &   \includegraphics[width=0.238\textwidth]{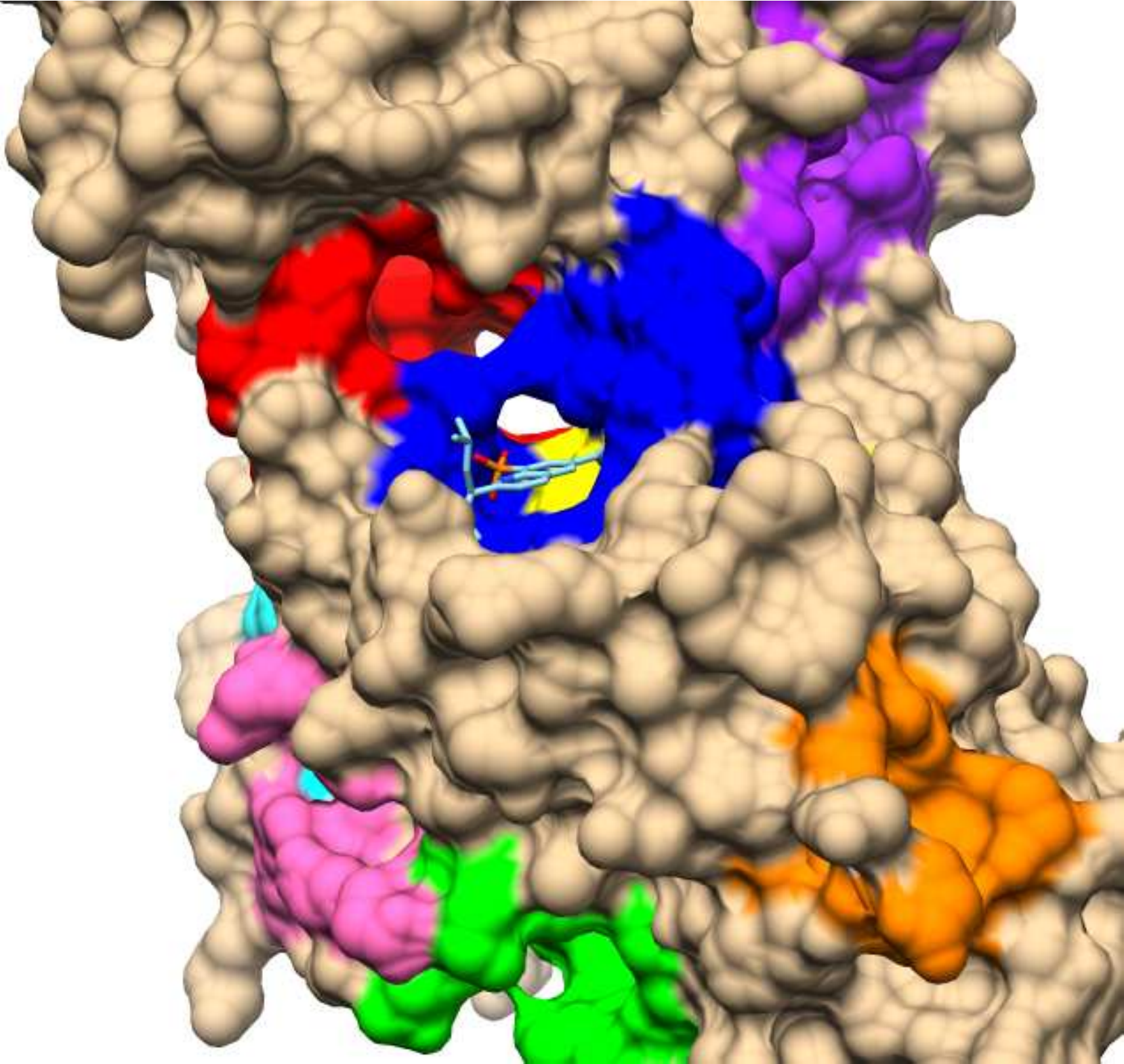} & 
    \includegraphics[width=0.2\textwidth]{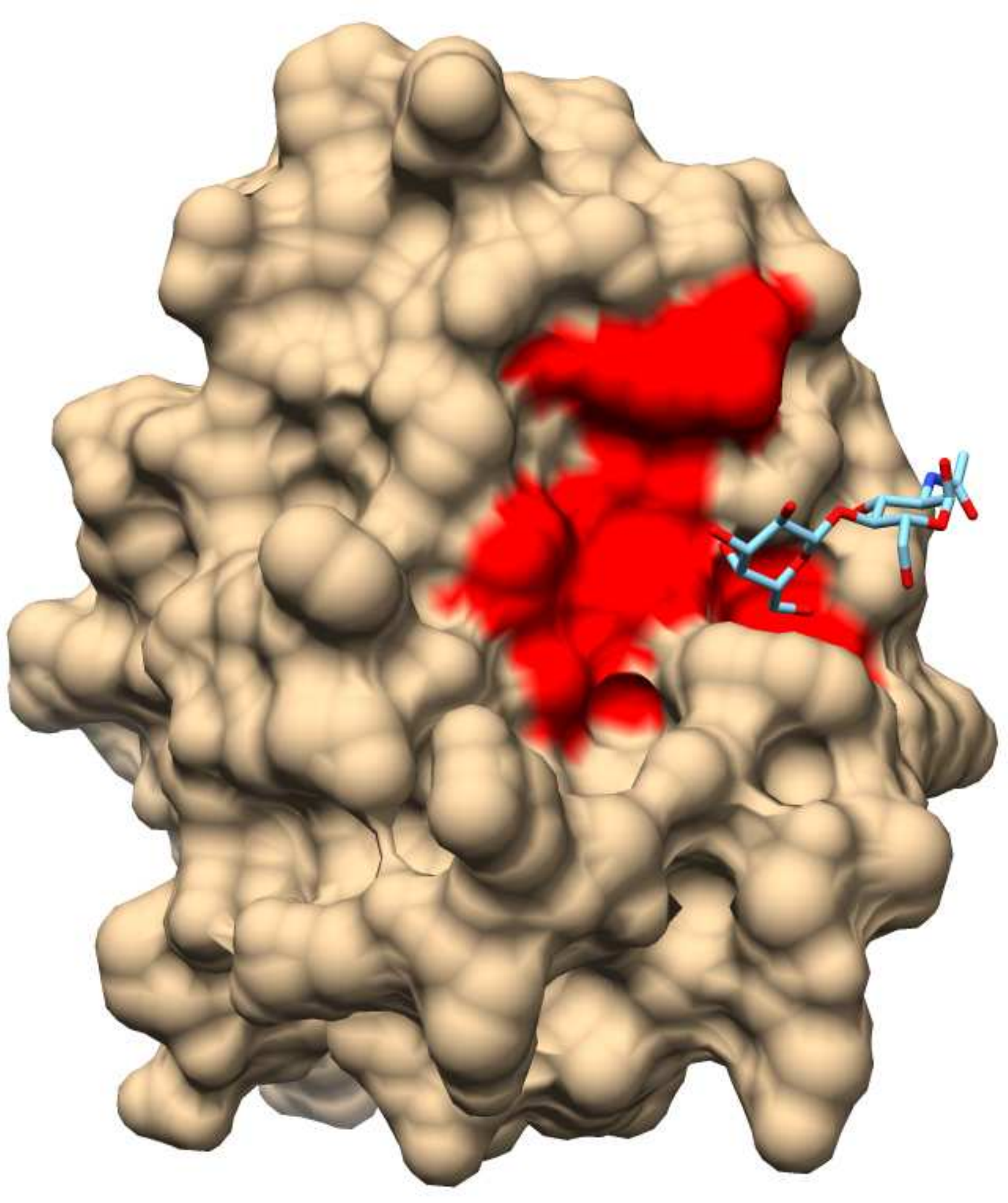} &   \includegraphics[width=0.2\textwidth]{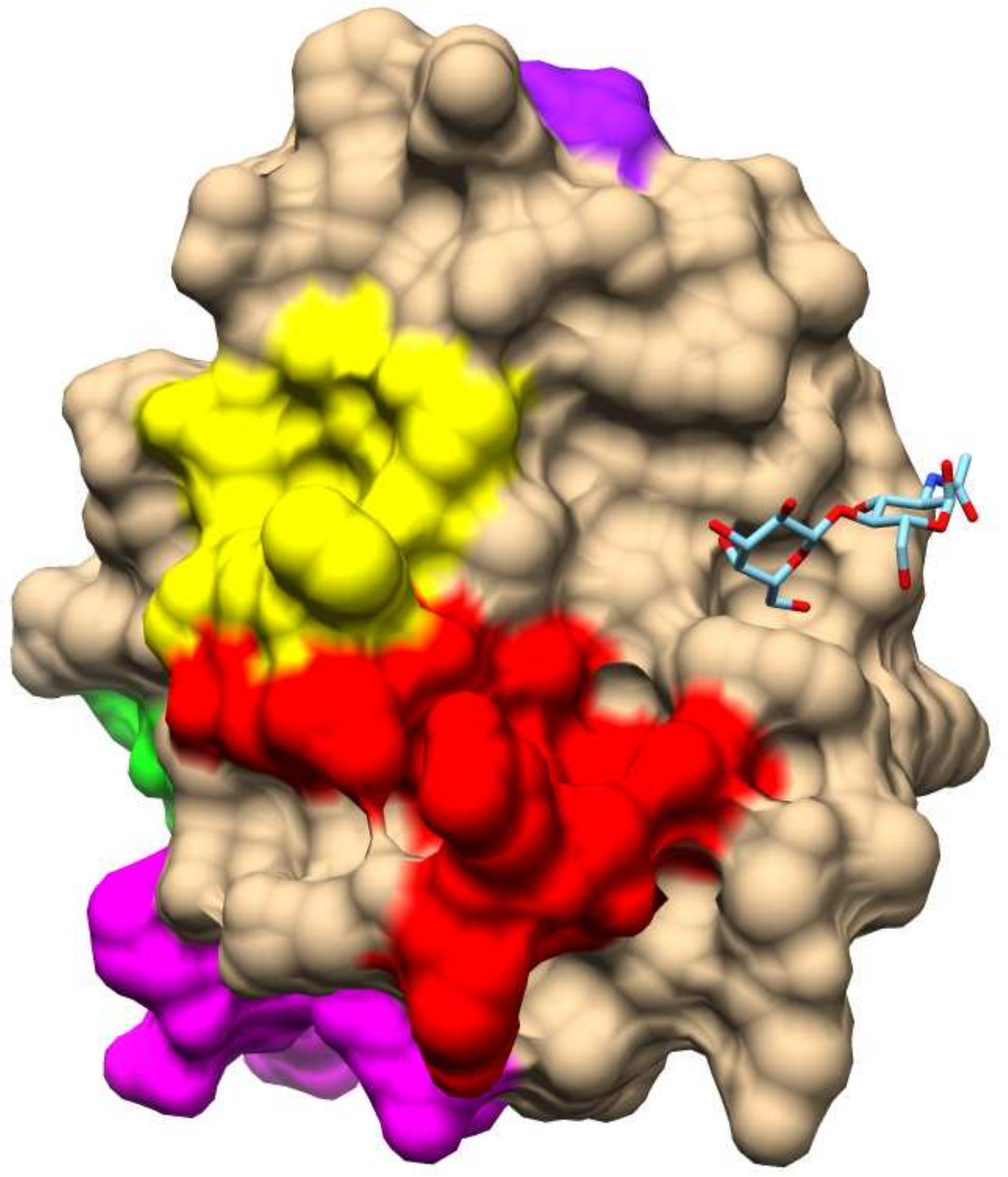}\\
    \qquad (a) & \qquad (b) & \quad (c) & \quad (d)\\
    \includegraphics[width=0.22\textwidth]{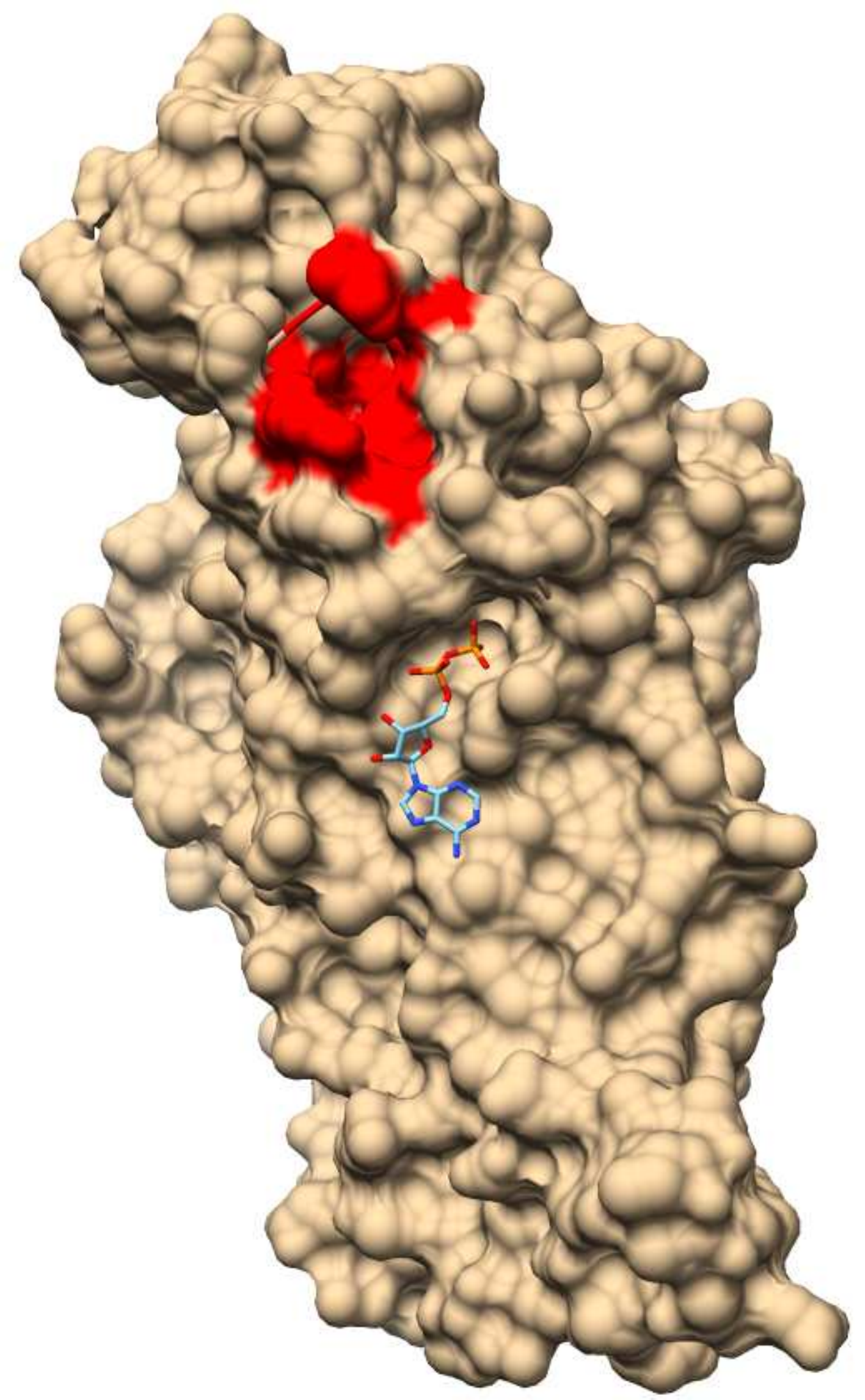} &   \includegraphics[width=0.22\textwidth]{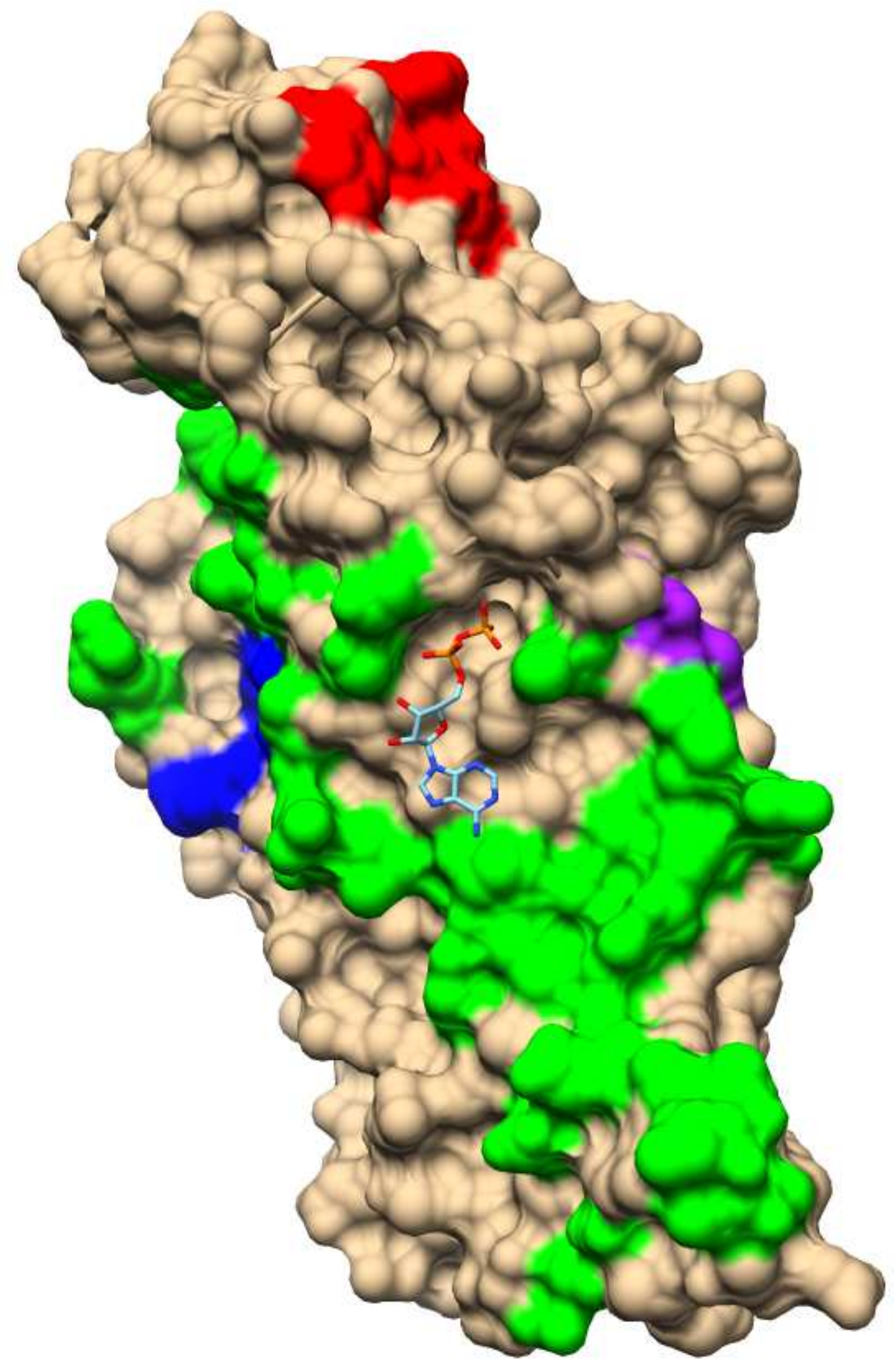} & 
    \includegraphics[width=0.22\textwidth]{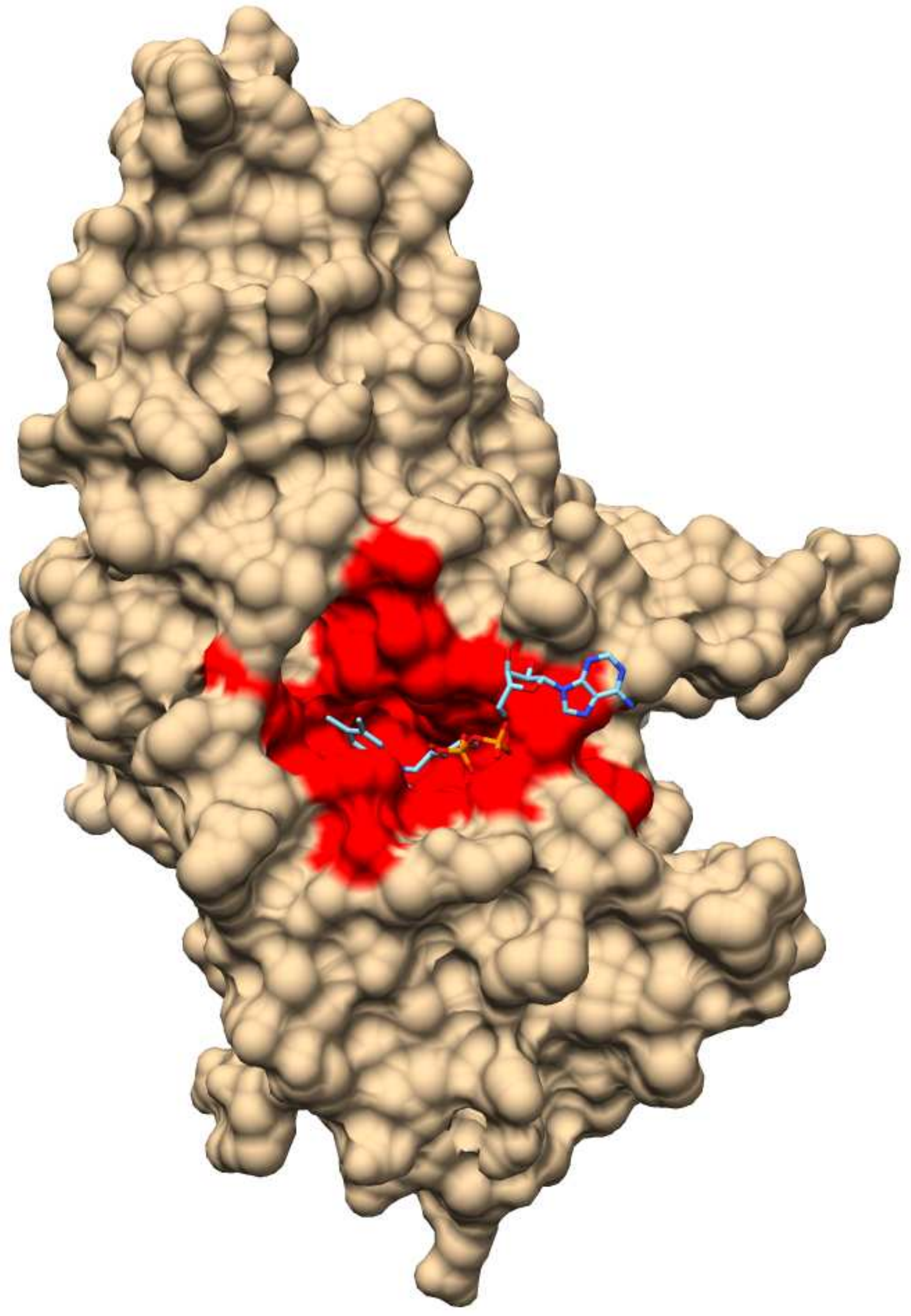} &   \includegraphics[width=0.22\textwidth]{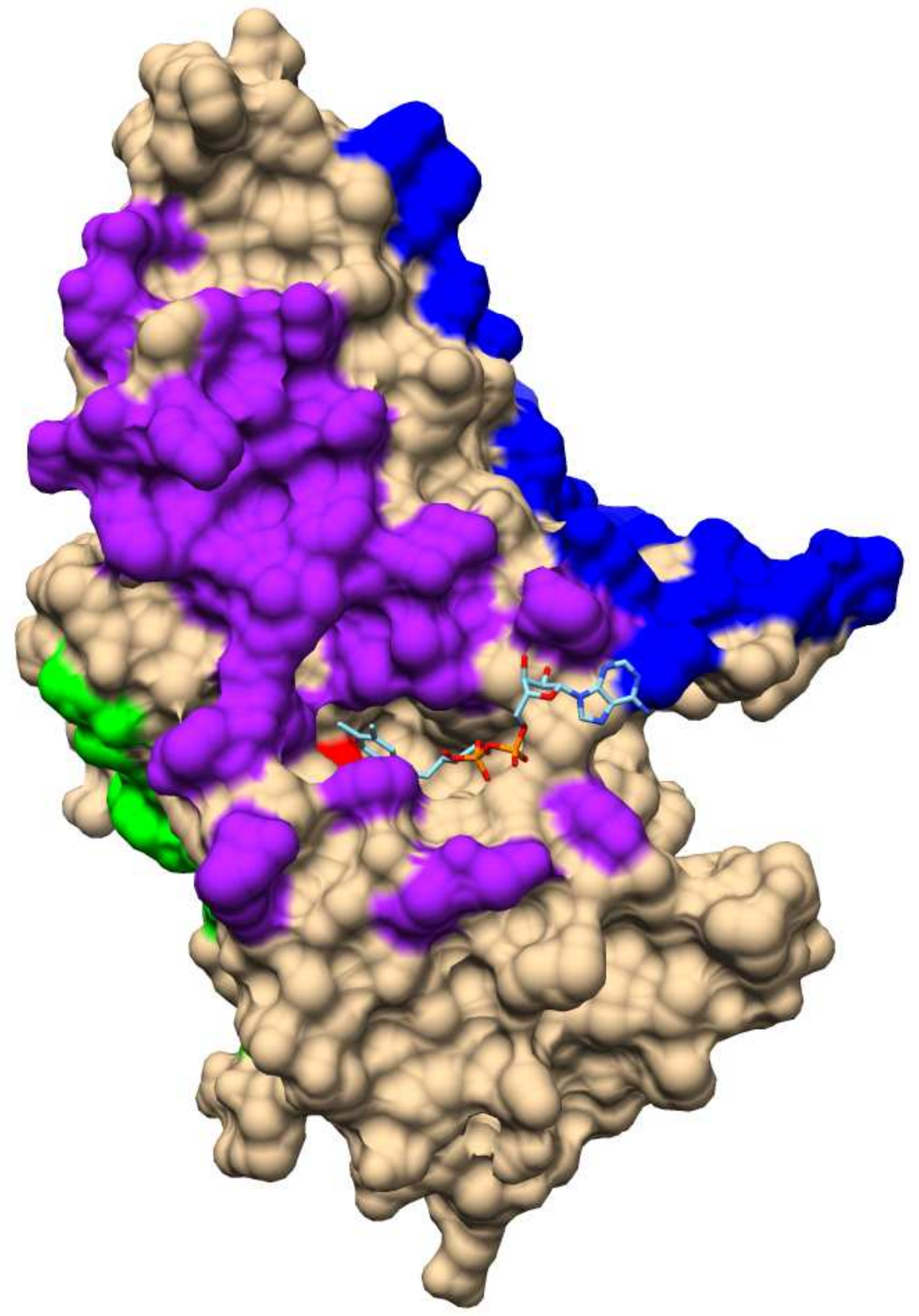}\\
    \qquad (e) & \qquad (f) & \quad (g) & \quad (h)
\end{tabular}
\caption[]{Extraction case studies by different methods on various structures. (a) DeepSurf on '1q6t', (b) AutoSite on '1q6t', (c) DeepSurf on '5galB', (d) AutoSite on '5galB', (e) DeepSurf on '3dsrA', (f) COFACTOR on '3dsrA', (g) DeepSurf on '2d29A', (h) COFACTOR on '2d29A'.} \label{fig:study_non_dl}
\end{figure*}

Some illustrative examples of the success or failure of each method in conjunction to the previous results are presented in Fig.~\ref{fig:study_non_dl}. One indicative case where the large number of extracted pockets affects the top-(n+2) accuracies is depicted in Fig.~\ref{fig:study_non_dl}(a) and (b) for DeepSurf and AutoSite respectively. Both methods extract approximately the same top solution (red) and fail to identify the real one for a small margin. Since DeepSurf extracts only one solution, it leads to a failure in both top-n and top-(n+2) accuracies. On the other hand, AutoSite extracts much more solutions (28 in total) and is able to locate it properly with the 2nd solution (blue), contributing thus to a succeeded prediction in top-(n+2) accuracy. Yet, a large number of extracted pockets does not always ensure a correct prediction. Fig.~\ref{fig:study_non_dl}(c) and (d) display the case of '5galB', where AutoSite fails to correctly locate the binding site despite the 6 extracted pockets, while DeepSurf identify it, even marginally, with one extraction. Figures~\ref{fig:study_non_dl}(e)-(h) examine the behavior of COFACTOR compared to DeepSurf. In case of '3dsrA', we notice that both methods fail in practice to correctly recognize the pocket. Nevertheless, the center of the 3rd solution of COFACTOR (green) falls accidentally close to the real pocket and is therefore counted as a top-(n+2) success, despite covering a wide area with no physical meaning. This case highlights the need for visual inspection of the results instead of solely depending on distance measures. A similar behavior by COFACTOR is also observed in Fig.~\ref{fig:study_non_dl}, although this time the extraction is not considered successful.

\section{Conclusion}

In this paper, a novel method, called DeepSurf, was presented for predicting potential druggable sites on proteins. The identification of promising candidate areas for binding on a protein's surface plays an important role towards drug discovery. DeepSurf proposes a novel approach on this task by combining a surface representation of the protein with a set of local 3D voxelized grids placed on protein's surface. After computing appropriate chemical features, these grids are iteratively imported to a state-of-the-art 3D convolutional network and the resulted ligandability scores of each surface point are, finally, clustered to create the binding sites. 

After comparing the proposed method with a set of competing deep learning methods in three diverse datasets of holo and apo structures, DeepSurf proved quite effective by outperforming the competing methods in terms of both localization and overlapping accuracies. Despite its domination on overlapping accuracies, the attained values remained quite small proving that the extraction of ideally shaped binding sites still remains an open issue. DeepSurf attained also competitive results when compared to a number of non-data-driven methods, although their tendency to extract a large number of pockets favored them when more solutions are considered. Finally, a sensitivity analysis on ligandability threshold showed that DeepSurf needs to preserve only the most reliably assigned by the network points in order to acquire its maximum performance.

\section*{Funding}

The work has been supported by the ATXN1-MED15 PPI project funded by the GSRT - Hellenic Foundation for Research and Innovation.

\bibliographystyle{unsrtnat}

\bibliography{document}

\begin{thebibliography}{37}
\providecommand{\natexlab}[1]{#1}
\providecommand{\url}[1]{\texttt{#1}}
\expandafter\ifx\csname urlstyle\endcsname\relax
  \providecommand{\doi}[1]{doi: #1}\else
  \providecommand{\doi}{doi: \begingroup \urlstyle{rm}\Url}\fi

\bibitem[Macari et~al.(2019)Macari, Toti, and
  Polticelli]{macari2019computational}
Gabriele Macari, Daniele Toti, and Fabio Polticelli.
\newblock Computational methods and tools for binding site recognition between
  proteins and small molecules: from classical geometrical approaches to modern
  machine learning strategies.
\newblock \emph{Journal of Computer-Aided Molecular Design}, 33\penalty0
  (10):\penalty0 887--903, 2019.

\bibitem[Capra et~al.(2009)Capra, Laskowski, Thornton, Singh, and
  Funkhouser]{capra2009predicting}
John~A Capra, Roman~A Laskowski, Janet~M Thornton, Mona Singh, and Thomas~A
  Funkhouser.
\newblock Predicting protein ligand binding sites by combining evolutionary
  sequence conservation and 3d structure.
\newblock \emph{PLoS computational biology}, 5\penalty0 (12):\penalty0
  e1000585, 2009.

\bibitem[Le~Guilloux et~al.(2009)Le~Guilloux, Schmidtke, and
  Tuffery]{le2009fpocket}
Vincent Le~Guilloux, Peter Schmidtke, and Pierre Tuffery.
\newblock Fpocket: an open source platform for ligand pocket detection.
\newblock \emph{BMC bioinformatics}, 10\penalty0 (1):\penalty0 168, 2009.

\bibitem[Dias et~al.(2017)Dias, Nguyen, Jorge, and Gomes]{dias2017multi}
S{\'e}rgio~ED Dias, Quoc~T Nguyen, Joaquim~A Jorge, and Abel~JP Gomes.
\newblock Multi-gpu-based detection of protein cavities using critical points.
\newblock \emph{Future Generation Computer Systems}, 67:\penalty0 430--440,
  2017.

\bibitem[Ngan et~al.(2011)Ngan, Hall, Zerbe, Grove, Kozakov, and
  Vajda]{ngan2011ftsite}
Chi-Ho Ngan, David~R Hall, Brandon Zerbe, Laurie~E Grove, Dima Kozakov, and
  Sandor Vajda.
\newblock Ftsite: high accuracy detection of ligand binding sites on unbound
  protein structures.
\newblock \emph{Bioinformatics}, 28\penalty0 (2):\penalty0 286--287, 2011.

\bibitem[Ravindranath and Sanner(2016)]{ravindranath2016autosite}
Pradeep~Anand Ravindranath and Michel~F Sanner.
\newblock Autosite: an automated approach for pseudo-ligands prediction—from
  ligand-binding sites identification to predicting key ligand atoms.
\newblock \emph{Bioinformatics}, 32\penalty0 (20):\penalty0 3142--3149, 2016.

\bibitem[Tsujikawa et~al.(2016)Tsujikawa, Sato, Wei, Saad, Sumikoshi, Nakamura,
  Terada, and Shimizu]{tsujikawa2016development}
Hiroto Tsujikawa, Kenta Sato, Cao Wei, Gul Saad, Kazuya Sumikoshi, Shugo
  Nakamura, Tohru Terada, and Kentaro Shimizu.
\newblock Development of a protein--ligand-binding site prediction method based
  on interaction energy and sequence conservation.
\newblock \emph{Journal of structural and functional genomics}, 17\penalty0
  (2-3):\penalty0 39--49, 2016.

\bibitem[Brylinski and Skolnick(2008)]{brylinski2008threading}
Michal Brylinski and Jeffrey Skolnick.
\newblock A threading-based method (findsite) for ligand-binding site
  prediction and functional annotation.
\newblock \emph{Proceedings of the National Academy of sciences}, 105\penalty0
  (1):\penalty0 129--134, 2008.

\bibitem[Hwang et~al.(2017)Hwang, Dey, Petrey, and Honig]{hwang2017structure}
Howook Hwang, Fabian Dey, Donald Petrey, and Barry Honig.
\newblock Structure-based prediction of ligand--protein interactions on a
  genome-wide scale.
\newblock \emph{Proceedings of the National Academy of Sciences}, 114\penalty0
  (52):\penalty0 13685--13690, 2017.

\bibitem[Toti et~al.(2017)Toti, Viet~Hung, Tortosa, Brandi, and
  Polticelli]{toti2017libra}
Daniele Toti, Le~Viet~Hung, Valentina Tortosa, Valentina Brandi, and Fabio
  Polticelli.
\newblock Libra-wa: a web application for ligand binding site detection and
  protein function recognition.
\newblock \emph{Bioinformatics}, 34\penalty0 (5):\penalty0 878--880, 2017.

\bibitem[Zhang et~al.(2011)Zhang, Li, Lin, Schroeder, and
  Huang]{zhang2011identification}
Zengming Zhang, Yu~Li, Biaoyang Lin, Michael Schroeder, and Bingding Huang.
\newblock Identification of cavities on protein surface using multiple
  computational approaches for drug binding site prediction.
\newblock \emph{Bioinformatics}, 27\penalty0 (15):\penalty0 2083--2088, 2011.

\bibitem[Yang et~al.(2013)Yang, Roy, and Zhang]{yang2013protein}
Jianyi Yang, Ambrish Roy, and Yang Zhang.
\newblock Protein--ligand binding site recognition using complementary
  binding-specific substructure comparison and sequence profile alignment.
\newblock \emph{Bioinformatics}, 29\penalty0 (20):\penalty0 2588--2595, 2013.

\bibitem[Jian et~al.(2016)Jian, Elumalai, Pitti, Wu, Tsai, Chang, Peng, and
  Yang]{jian2016predicting}
Jhih-Wei Jian, Pavadai Elumalai, Thejkiran Pitti, Chih~Yuan Wu, Keng-Chang
  Tsai, Jeng-Yih Chang, Hung-Pin Peng, and An-Suei Yang.
\newblock Predicting ligand binding sites on protein surfaces by 3-dimensional
  probability density distributions of interacting atoms.
\newblock \emph{PloS one}, 11\penalty0 (8):\penalty0 e0160315, 2016.

\bibitem[Kriv{\'a}k and Hoksza(2018)]{krivak2018p2rank}
Radoslav Kriv{\'a}k and David Hoksza.
\newblock P2rank: machine learning based tool for rapid and accurate prediction
  of ligand binding sites from protein structure.
\newblock \emph{Journal of cheminformatics}, 10\penalty0 (1):\penalty0 39,
  2018.

\bibitem[Ragoza et~al.(2017)Ragoza, Hochuli, Idrobo, Sunseri, and
  Koes]{ragoza2017protein}
Matthew Ragoza, Joshua Hochuli, Elisa Idrobo, Jocelyn Sunseri, and David~Ryan
  Koes.
\newblock Protein--ligand scoring with convolutional neural networks.
\newblock \emph{Journal of chemical information and modeling}, 57\penalty0
  (4):\penalty0 942--957, 2017.

\bibitem[Imrie et~al.(2018)Imrie, Bradley, van~der Schaar, and
  Deane]{imrie2018protein}
Fergus Imrie, Anthony~R Bradley, Mihaela van~der Schaar, and Charlotte~M Deane.
\newblock Protein family-specific models using deep neural networks and
  transfer learning improve virtual screening and highlight the need for more
  data.
\newblock \emph{Journal of chemical information and modeling}, 58\penalty0
  (11):\penalty0 2319--2330, 2018.

\bibitem[Stepniewska-Dziubinska et~al.(2018)Stepniewska-Dziubinska,
  Zielenkiewicz, and Siedlecki]{stepniewska2018development}
Marta~M Stepniewska-Dziubinska, Piotr Zielenkiewicz, and Pawel Siedlecki.
\newblock Development and evaluation of a deep learning model for
  protein--ligand binding affinity prediction.
\newblock \emph{Bioinformatics}, 34\penalty0 (21):\penalty0 3666--3674, 2018.

\bibitem[Jim{\'e}nez et~al.(2018)Jim{\'e}nez, Skalic, Martinez-Rosell, and
  De~Fabritiis]{jimenez2018k}
Jos{\'e} Jim{\'e}nez, Miha Skalic, Gerard Martinez-Rosell, and Gianni
  De~Fabritiis.
\newblock K deep: protein--ligand absolute binding affinity prediction via
  3d-convolutional neural networks.
\newblock \emph{Journal of chemical information and modeling}, 58\penalty0
  (2):\penalty0 287--296, 2018.

\bibitem[Wang et~al.(2016)Wang, Peng, Ma, and Xu]{wang2016protein}
Sheng Wang, Jian Peng, Jianzhu Ma, and Jinbo Xu.
\newblock Protein secondary structure prediction using deep convolutional
  neural fields.
\newblock \emph{Scientific reports}, 6:\penalty0 18962, 2016.

\bibitem[Senior et~al.(2019)Senior, Evans, Jumper, Kirkpatrick, Sifre, Green,
  Qin, {\v{Z}}{\'\i}dek, Nelson, Bridgland, et~al.]{senior2019protein}
Andrew~W Senior, Richard Evans, John Jumper, James Kirkpatrick, Laurent Sifre,
  Tim Green, Chongli Qin, Augustin {\v{Z}}{\'\i}dek, Alexander~WR Nelson, Alex
  Bridgland, et~al.
\newblock Protein structure prediction using multiple deep neural networks in
  the 13th critical assessment of protein structure prediction (casp13).
\newblock \emph{Proteins: Structure, Function, and Bioinformatics}, 87\penalty0
  (12):\penalty0 1141--1148, 2019.

\bibitem[Jim{\'e}nez et~al.(2017)Jim{\'e}nez, Doerr, Mart{\'\i}nez-Rosell,
  Rose, and De~Fabritiis]{jimenez2017deepsite}
Jos{\'e} Jim{\'e}nez, Stefan Doerr, Gerard Mart{\'\i}nez-Rosell, Alexander~S
  Rose, and Gianni De~Fabritiis.
\newblock Deepsite: protein-binding site predictor using 3d-convolutional
  neural networks.
\newblock \emph{Bioinformatics}, 33\penalty0 (19):\penalty0 3036--3042, 2017.

\bibitem[Jiang et~al.(2019{\natexlab{a}})Jiang, Li, Bian, and
  Wei]{jiang2019novel}
Mingjian Jiang, Zhen Li, Yujie Bian, and Zhiqiang Wei.
\newblock A novel protein descriptor for the prediction of drug binding sites.
\newblock \emph{BMC bioinformatics}, 20\penalty0 (1):\penalty0 1--13,
  2019{\natexlab{a}}.

\bibitem[Stepniewska-Dziubinska et~al.(2020)Stepniewska-Dziubinska,
  Zielenkiewicz, and Siedlecki]{stepniewska2020improving}
Marta~M Stepniewska-Dziubinska, Piotr Zielenkiewicz, and Pawel Siedlecki.
\newblock improving detection of protein-ligand binding sites with 3d
  segmentation.
\newblock \emph{Scientific reports}, 10\penalty0 (1):\penalty0 1--9, 2020.

\bibitem[Jiang et~al.(2019{\natexlab{b}})Jiang, Wei, Zhang, Wang, Wang, and
  Li]{jiang2019frsite}
Mingjian Jiang, Zhiqiang Wei, Shugang Zhang, Shuang Wang, Xiaofeng Wang, and
  Zhen Li.
\newblock Frsite: Protein drug binding site prediction based on faster r--cnn.
\newblock \emph{Journal of Molecular Graphics and Modelling}, 93:\penalty0
  107454, 2019{\natexlab{b}}.

\bibitem[Axenopoulos et~al.(2015)Axenopoulos, Rafailidis, Papadopoulos,
  Houstis, and Daras]{axenopoulos2015similarity}
Apostolos Axenopoulos, Dimitrios Rafailidis, Georgios Papadopoulos, Elias~N
  Houstis, and Petros Daras.
\newblock Similarity search of flexible 3d molecules combining local and global
  shape descriptors.
\newblock \emph{IEEE/ACM transactions on computational biology and
  bioinformatics}, 13\penalty0 (5):\penalty0 954--970, 2015.

\bibitem[Comaniciu and Meer(2002)]{comaniciu2002mean}
Dorin Comaniciu and Peter Meer.
\newblock Mean shift: A robust approach toward feature space analysis.
\newblock \emph{IEEE Transactions on Pattern Analysis \& Machine Intelligence},
  24\penalty0 (5):\penalty0 603--619, 2002.

\bibitem[He et~al.(2016)He, Zhang, Ren, and Sun]{he2016deep}
Kaiming He, Xiangyu Zhang, Shaoqing Ren, and Jian Sun.
\newblock Deep residual learning for image recognition.
\newblock In \emph{Proceedings of the IEEE conference on computer vision and
  pattern recognition}, pages 770--778, 2016.

\bibitem[Dimou et~al.(2018)Dimou, Ataloglou, Dimitropoulos, Alvarez, and
  Daras]{dimou2018lds}
Anastasios Dimou, Dimitrios Ataloglou, Kosmas Dimitropoulos, Federico Alvarez,
  and Petros Daras.
\newblock Lds-inspired residual networks.
\newblock \emph{IEEE Transactions on Circuits and Systems for Video
  Technology}, 2018.

\bibitem[Desaphy et~al.(2014)Desaphy, Bret, Rognan, and
  Kellenberger]{desaphy2014sc}
J{\'e}r{\'e}my Desaphy, Guillaume Bret, Didier Rognan, and Esther Kellenberger.
\newblock sc-pdb: a 3d-database of ligandable binding sites—10 years on.
\newblock \emph{Nucleic acids research}, 43\penalty0 (D1):\penalty0 D399--D404,
  2014.

\bibitem[Schmidtke et~al.(2010)Schmidtke, Souaille, Estienne, Baurin, and
  Kroemer]{schmidtke2010large}
Peter Schmidtke, Catherine Souaille, Fr{\'e}d{\'e}ric Estienne, Nicolas Baurin,
  and Romano~T Kroemer.
\newblock Large-scale comparison of four binding site detection algorithms.
\newblock \emph{Journal of chemical information and modeling}, 50\penalty0
  (12):\penalty0 2191--2200, 2010.

\bibitem[Chen et~al.(2011)Chen, Mizianty, Gao, and Kurgan]{chen2011critical}
Ke~Chen, Marcin~J Mizianty, Jianzhao Gao, and Lukasz Kurgan.
\newblock A critical comparative assessment of predictions of protein-binding
  sites for biologically relevant organic compounds.
\newblock \emph{Structure}, 19\penalty0 (5):\penalty0 613--621, 2011.

\bibitem[UCSF(2020)]{dms}
UCSF.
\newblock Dms software.
\newblock
  \url{http://www.cgl.ucsf.edu/chimera/docs/UsersGuide/midas/dms1.html}, 2020.
\newblock Accessed: 2020-08-30.

\bibitem[Johnson and Khoshgoftaar(2019)]{johnson2019survey}
Justin~M Johnson and Taghi~M Khoshgoftaar.
\newblock Survey on deep learning with class imbalance.
\newblock \emph{Journal of Big Data}, 6\penalty0 (1):\penalty0 27, 2019.

\bibitem[Kingma and Ba(2014)]{kingma2014adam}
Diederik~P Kingma and Jimmy Ba.
\newblock Adam: A method for stochastic optimization.
\newblock \emph{arXiv preprint arXiv:1412.6980}, 2014.

\bibitem[Luscombe et~al.(2001)Luscombe, Laskowski, and
  Thornton]{luscombe2001amino}
Nicholas~M Luscombe, Roman~A Laskowski, and Janet~M Thornton.
\newblock Amino acid--base interactions: a three-dimensional analysis of
  protein--dna interactions at an atomic level.
\newblock \emph{Nucleic acids research}, 29\penalty0 (13):\penalty0 2860--2874,
  2001.

\bibitem[Cimermancic et~al.(2016)Cimermancic, Weinkam, Rettenmaier, Bichmann,
  Keedy, Woldeyes, Schneidman-Duhovny, Demerdash, Mitchell, Wells,
  et~al.]{cimermancic2016cryptosite}
Peter Cimermancic, Patrick Weinkam, T~Justin Rettenmaier, Leon Bichmann,
  Daniel~A Keedy, Rahel~A Woldeyes, Dina Schneidman-Duhovny, Omar~N Demerdash,
  Julie~C Mitchell, James~A Wells, et~al.
\newblock Cryptosite: expanding the druggable proteome by characterization and
  prediction of cryptic binding sites.
\newblock \emph{Journal of molecular biology}, 428\penalty0 (4):\penalty0
  709--719, 2016.

\bibitem[Roy et~al.(2012)Roy, Yang, and Zhang]{roy2012cofactor}
Ambrish Roy, Jianyi Yang, and Yang Zhang.
\newblock Cofactor: an accurate comparative algorithm for structure-based
  protein function annotation.
\newblock \emph{Nucleic acids research}, 40\penalty0 (W1):\penalty0 W471--W477,
  2012.

\end{thebibliography}

\newpage

\setcounter{figure}{0}
\renewcommand{\thefigure}{S\arabic{figure}}

%\section*{Supplementary Figures}
%\textbf{Supplementary Figures}

\begin{figure*}[t]\centering
\begin{tabular}{c}
    \includegraphics[width=0.8\textwidth]{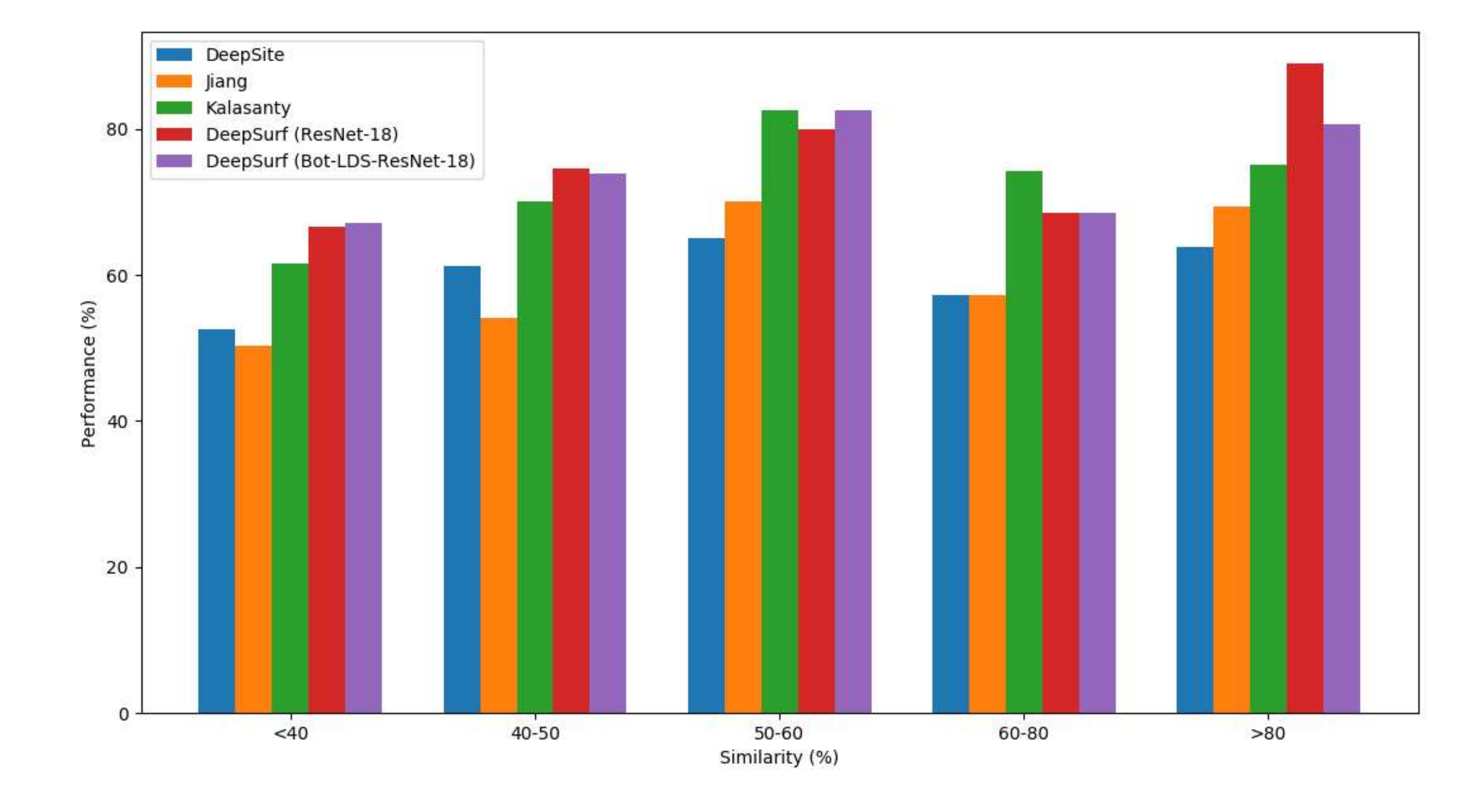}\\
    \qquad (a) COACH420\\[10pt]
    \includegraphics[width=0.8\textwidth]{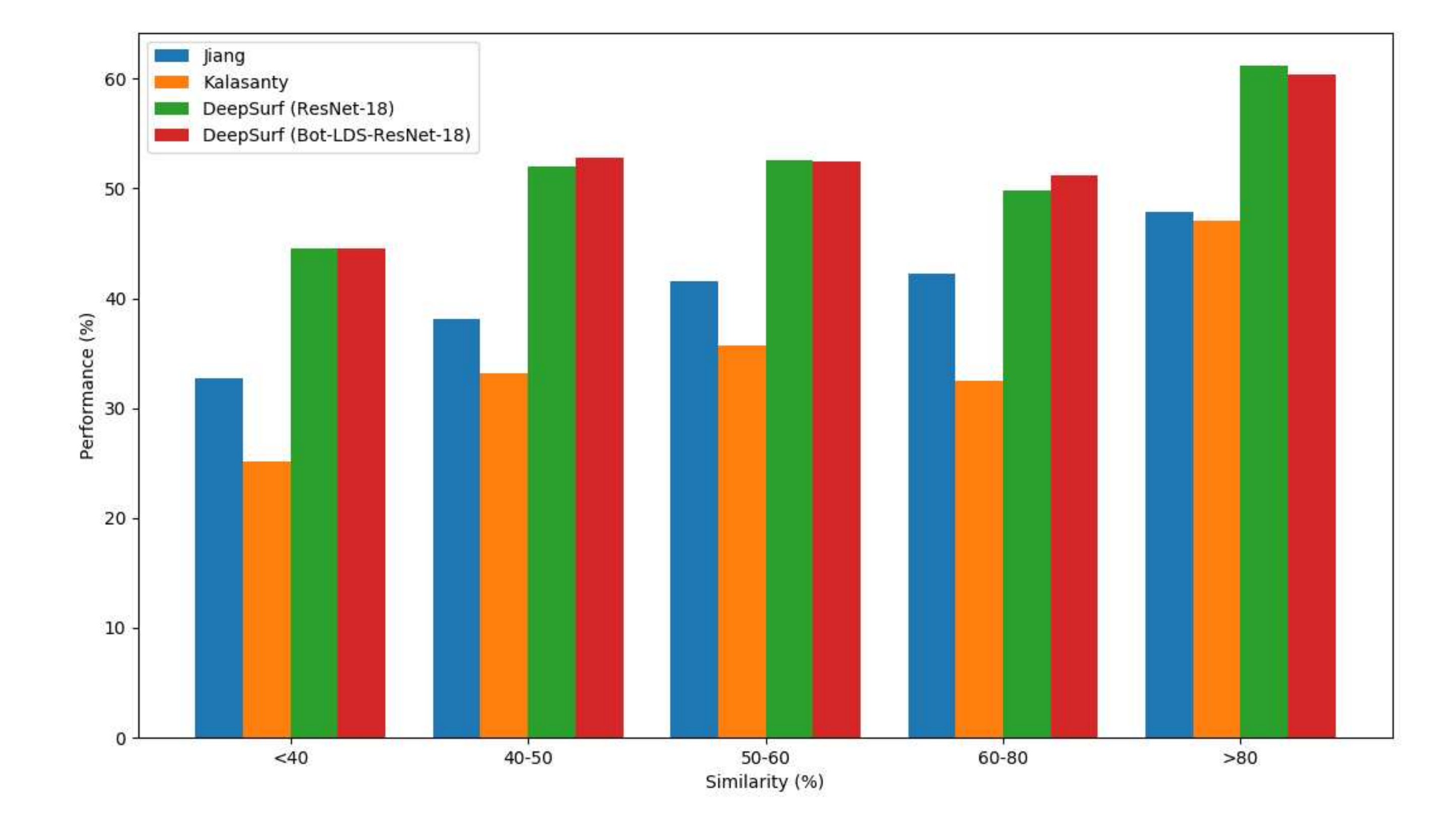} \\
    \qquad (b) HOLO4K
\end{tabular}
\caption[Performance vs similarity for DL-driven methods (COACH420 -  HOLO4K)]{Top-n performance (\%) of DeepSurf and competing DL-based methods using the DCA criterion ($D_{cut}=4$ \angstrom) for various ranges of global sequence similarity between train and test sets for (a) COACH420 and (b) HOLO4K. DeepSite results were obtained from Krivák and Hoksza (2018), and we were unable to retrieve detailed per-protein results for the HOLO4K case.} 
\end{figure*}

\begin{figure*}[!t]\centering
\begin{tabular}{cc}
    \includegraphics[width=0.50\textwidth]{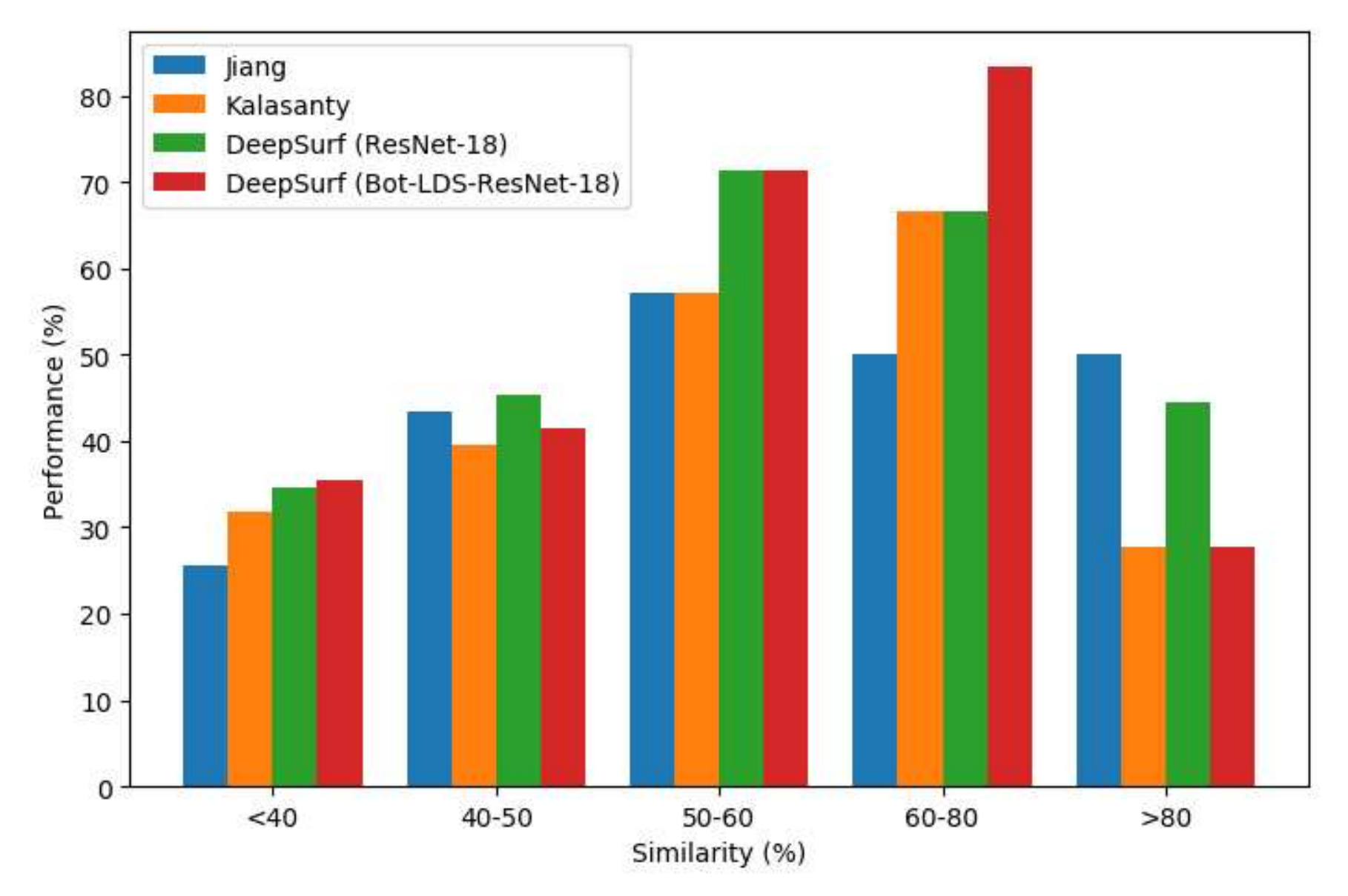} &
    \includegraphics[width=0.45\textwidth]{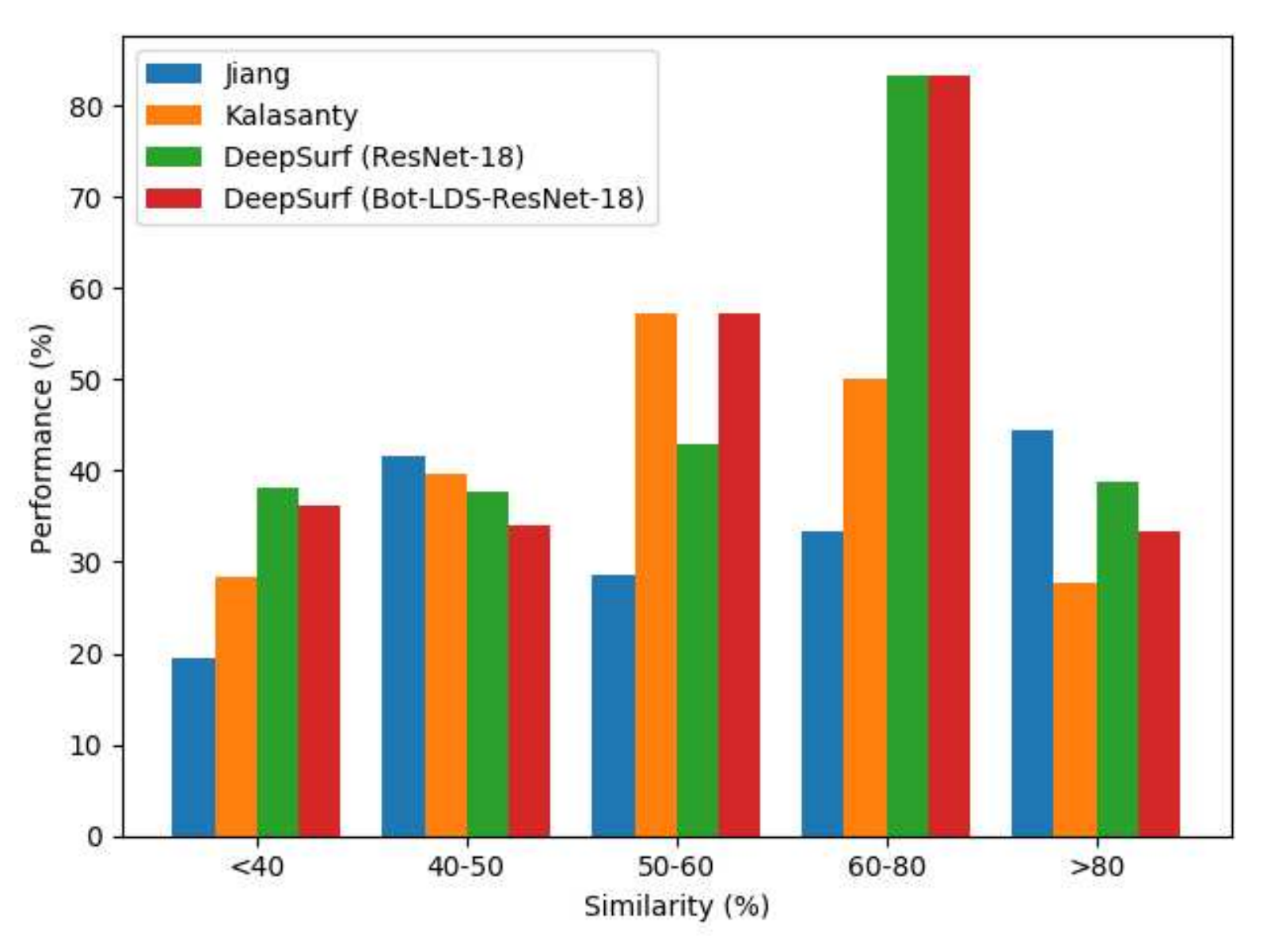} \\
    \quad (a) CHEN\_holo & \quad (b) CHEN\_apo
\end{tabular}
\caption[Performance vs similarity for DL-driven methods (CHEN\_holo - CHEN\_apo)]{Top-n performance (\%) of DeepSurf and competing DL-based methods using the DCA criterion ($D_{cut}=4$ \angstrom) for various ranges of global sequence similarity between train and test sets for (a) CHEN\_holo and (b) CHEN\_apo.}
\end{figure*}

\begin{figure*}[!t]\centering
\begin{tabular}{cc}
    \includegraphics[width=0.54\textwidth]{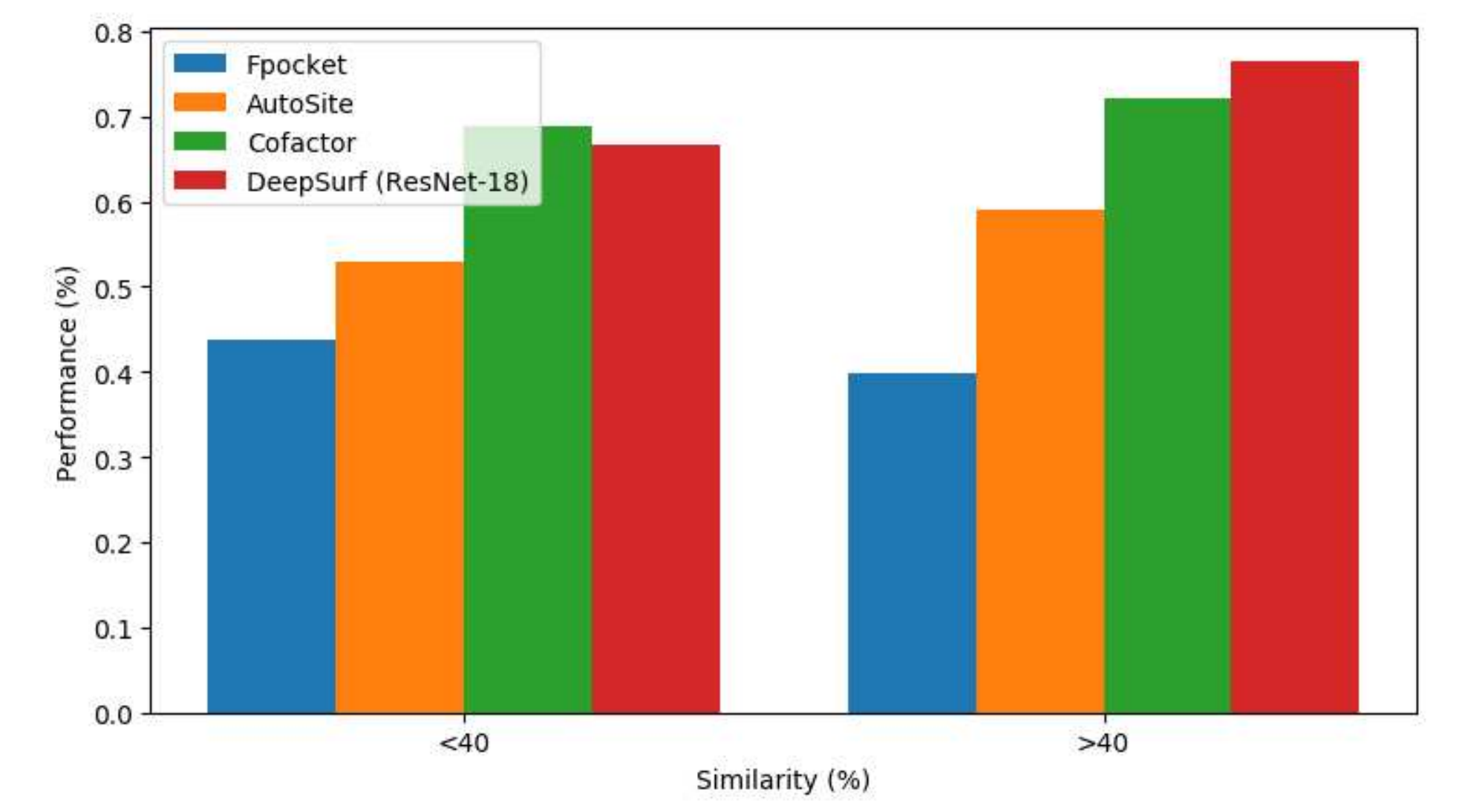} &
    \includegraphics[width=0.42\textwidth]{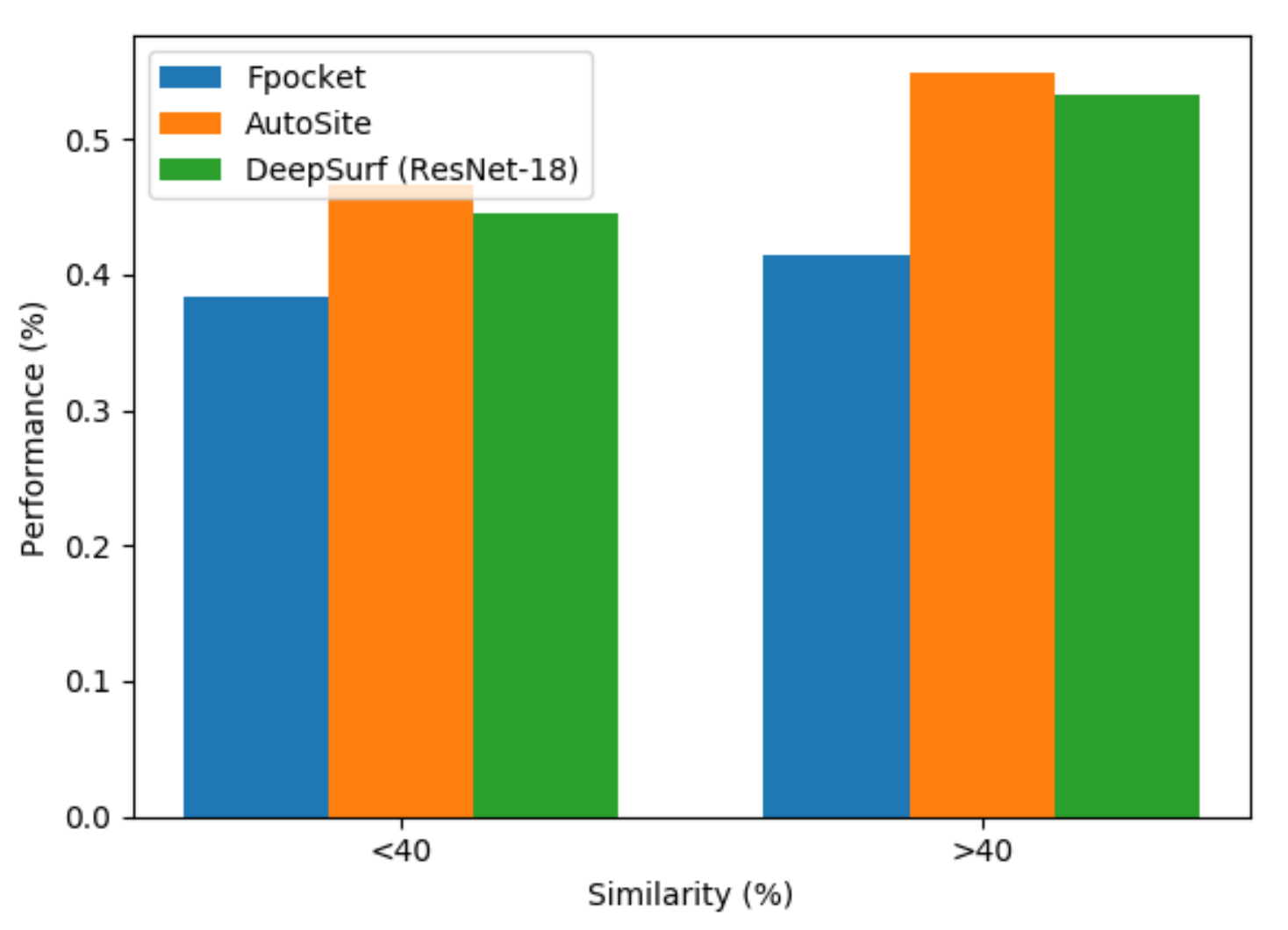} \\
    \qquad (a) COACH420 & \qquad (b) HOLO4K \\%[12pt]
\end{tabular}
\caption[Performance vs similarity for non-data-driven methods (COACH420 - HOLO4K)]{Top-n performance (\%) of DeepSurf and competing non-data-driven methods using the DCA criterion ($D_{cut}=4$ \angstrom) on the homologous ($>$40\%) and non-homologous ($<$40\%) subsets of (a) COACH420 and (b) HOLO4K. Due to high execution times, COFACTOR was tested only on the smaller COACH420 set.}
\end{figure*}

\end{document}